\documentclass[aps,pre,showpacs,showkeys,twocolumn,superscriptaddress,floats,floatfix]{revtex4-2}

\usepackage{graphicx}
\usepackage{float}
\usepackage{dcolumn}
\usepackage{bm}
\usepackage{subcaption}
\usepackage{caption}
\usepackage{comment}
\usepackage{soul}
\usepackage{ragged2e}
\usepackage{blindtext}
\usepackage{hyperref}
\usepackage{ulem}
\usepackage{amsmath}
\usepackage[dvipsnames]{xcolor}
\def \ii {{\rm i}}
\def \jj {{\rm j}}
\def \uu {{\bf u}}
\def \xx {{\rm \bf x}}
\def \rr {{\bf r}}
\def \vp {{\rm v_p}}
\def \phat {\hat{\bf p}}
\def \Fij {{\bf F}_{\rm ij}}
\def \etaT {{\boldsymbol \eta}^{\rm T}}
\def \etaR {\eta^{\rm R}}
\def \uv {\overline {{\hat {\bf u}} \cdot {\hat {\bf v}}}}
\def \vd {{\rm v}_0}
\def \OP {{\mathcal {O}}}
\def \Ttrans {T_{\rm trans}}

\begin{document}

\title{Effect of the background flow on the motility induced phase separation}

\author{Soni D. Prajapati}
 \affiliation{Department of Physics, Indian Institute of Technology Hyderabad, Hyderabad 502284, India}
 
\author{Akshay Bhatnagar}%
\email[Correspondence: ]{akshayphy@gmail.com}
\affiliation{Department of Physics, Indian Institute of Technology Palakkad,
Palakkad 678623, Kerala, India.}
\affiliation{Department of Physics, Indian Institute of Technology Roorkee, Roorkee 247667,
Uttarakhand, India}

\author{Anupam Gupta}
\email[Correspondence: ]{agupta@phy.iith.ac.in}
\affiliation{Department of Physics, Indian Institute of Technology Hyderabad, Hyderabad 502284, India}

\begin{abstract}
We simulate active Brownian particles (ABPs) with soft-repulsive interactions
subjected to a four-roll-mill flow. In the absence of flow, this system exhibits
motility-induced phase separation (MIPS). To investigate the interplay between
MIPS and flow-induced mixing, we introduce dimensionless parameters: a scaled
time, $\rm \tau$, and a scaled speed, ${\rm V}$, characterizing the ratio of ABP
to fluid time and speed scales, respectively. The parameter space defined by
$\rm \tau$ and ${\rm V}$ reveals three distinct ABP distribution regimes. At low speeds ${\rm
V} \ll 1$, flow dominates, leading to a homogeneous mixture. Conversely, at high speeds ${\rm V} \gg 1$, motility prevails, resulting in MIPS. In the
intermediate regime (${\rm V} \sim 1$), the system's behavior depends on $\rm \tau$.
For $\rm \tau <1$, a moderately mixed homogeneous phase emerges, while for $\rm \tau
>1$, a novel phenomenon, termed flow-induced phase separation (FIPS), arises due to
the combined effects of flow topology and ABP motility and size. To characterize
these phases, we analyze drift velocity, diffusivity, mean-squared displacement,
giant number fluctuations, radial distribution function, and cluster-size
distribution
\end{abstract}


\maketitle

\section{Introduction}
\label{sec:level1}

Active matter systems consist of entities that continuously consume energy from
their environment to induce internal changes, typically resulting in directed
motion \cite{marchetti2013hydrodynamics,bechinger2016active}. These systems
represent a subset of non-equilibrium systems observed across various length
scales in nature, from the microscopic (e.g., bacteria
\cite{wu2009periodic,sokolov2010swimming}, sperm cells) to the macroscopic
(e.g., fish \cite{crosato2018informative}, birds
\cite{nagy2010hierarchical,mora2016local}, animals
\cite{hueschen2023wildebeest,garcimartin2015flow}, and humans
\cite{bellomo2016human}). To gain deeper insights into these systems,
researchers have developed and studied numerous synthetic active particles in
controlled laboratory environments, including Janus particles
\cite{jiang2010active}, vibrating beds of self-propelling particles
\cite{Narayan2007}, active droplets \cite{dey2022oscillatory}, and robotic
systems \cite{o2017oscillators,paramanick2024programming}. Due to underlying
interactions among the constituent entities of active systems, these systems
exhibit emergent phenomena. Prominent examples include motility-induced phase
separation (MIPS)~\cite{redner2013structure,Fily2012}, characterized by density
fluctuations arising from self-propulsion, flocking, or herding behavior
displaying collective polar order~\cite{vicsek1995novel}, and the emergence of
nematic (apolar) ordering due to intrinsic anisotropy in the system
components~\cite{Narayan2007}.

Active systems typically coexist with surrounding fluids, as exemplified by
bacterial colonies~\cite{wu2009periodic,sokolov2010swimming}, active
colloids~\cite{jiang2010active}, and fish schools~\cite{crosato2018informative}.
However, the influence of the underlying flow on the emergent behavior of
these systems remains largely unexplored. This study aims to elucidate how
background flow affects the formation and characteristics of MIPS.

MIPS can be observed in various active particle models, including Active
Ornstein-Uhlenbeck particles~\cite{martin2021aoup}, Active Brownian Particles
(ABPs)~\cite{Fily2012,redner2013structure,sanoria2021influence}, and run-and-tumble
particles~\cite{bertrand2018optimized,santra2020run}. Despite model variations,
these systems exhibit qualitatively similar phase separation behavior. In this
study, we employ ABPs with soft repulsion and no alignment to simulate
microorganisms in a fluid environment. System dynamics in the absence of flow
agrees with previous findings~\cite{Fily2012,redner2013structure}. We are always
in the parameter regime where we observe MIPS without background
flow. MIPS displays distinct characteristics compared to equilibrium phase
separation, including reduced diffusivity and drift~\cite{Fily2012}, giant
number fluctuations (with standard deviation $\rm \Delta N$ scaling nearly linearly
with mean particle number $\rm N$)~\cite{toner2005hydrodynamics,Narayan2007},
ordered packing in the dense phase~\cite{sanoria2022percolation}, and a bimodal cluster size
distribution~\cite{sanoria2022percolation,dolai2018phase}. This work investigates how these features
are modified by the introduction of background flow.

\begin{figure}[!htb]
  \includegraphics[width=0.90\linewidth]{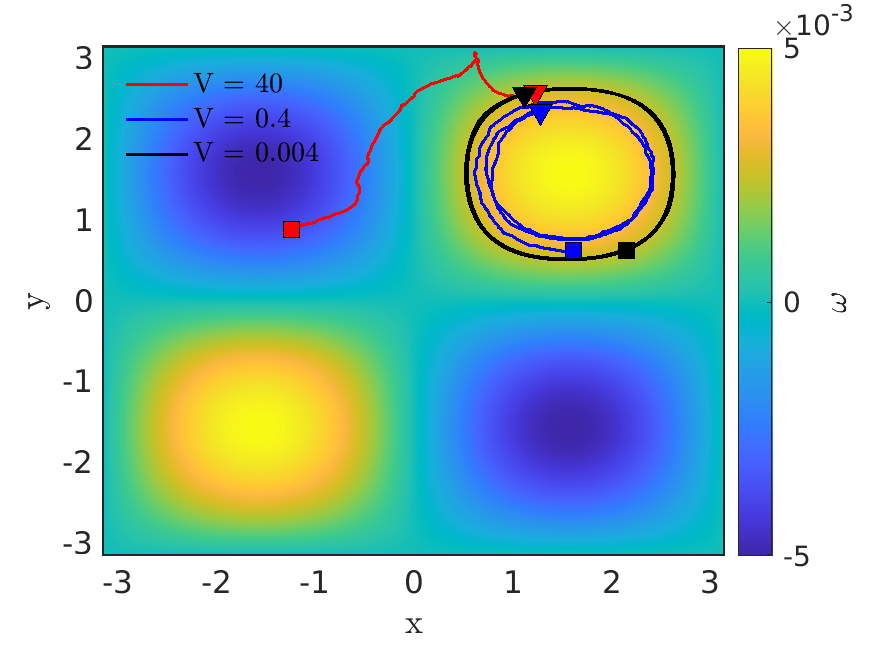}
\caption{\label{fig:Traj}\justifying{Representative Trajectories of ABPs with $\rm \tau = 400$ and $\rm \phi =0.7$ for three scaled-speeds overlaid on the vorticity field corresponding to four-roll-mill flow. The triangle and the square represent the starting and ending points of the trajectories, respectively.}}
\end{figure}

Since this is the first attempt to address such questions, we begin by
considering a controlled flow, a four-roll-mill flow
~\cite{gutierrez2019proper,lee2007microfluidic} (Fig.~\ref{fig:Traj}). This flow
exhibits a clear distinction between vortical and strain-dominated regions.
Known for its efficiency in mixing passive scalars
\cite{solomon2003uniform,solomon1988chaotic}, this flow creates a competition
with the demixing tendency inherent to MIPS. To quantify the relative importance
of flow and particle motility, we introduce dimensionless parameters based on
speed and time scales and our analysis explores the
dynamics within this parameter space.

\section{Model and Simulations}
\label{section2}

We consider a collection of $\rm N$ interacting ABPs,
confined to two dimensions, and suspended in a fluid. The ABPs are modeled as
disks of radius $\rm{a}$, each propelled with a self-propulsion speed $\rm \vp$ along the
orientation vector $\phat_\ii$, where $\ii$ ranges from $\rm 1$ to $\rm N$. In two
dimensions, $\phat_\ii$ is expressed as $\rm (\cos{(\theta_\ii)},
\sin{(\theta_\ii)})$, with $\rm \theta_\ii$ denoting the angle between the
orientation vector of the $\ii$-th particle and the $\rm x$-axis. The equations
of motion, for the particle with position, $\rr_\ii$ are as follows:

\begin{subequations}
\begin{equation}\label{position_tem}
\rm \dot{\rr}_\ii =\rm \vp \phat_\ii + {\mu_{t}} \sum_{\jj \neq \ii} \Fij + \etaT_{\ii}(t) + \uu(\rr_{\ii},t), 
 \end{equation} 
 \begin{equation}\label{theta_tem}
 \rm \dot{\theta_\ii} =\rm \etaR_\ii(t) + \frac{1}{2}\omega(\rr_\ii,t),
 \end{equation}
\end{subequations}

where, $\Fij$ models the soft repulsive interaction between the $\ii$-th and
$\jj$-th particle and is given as $\rm \Fij = -k(2a -|r_{\ii \jj}|) \hat{\rr}_{\ii
\jj}$, if $\rm |r_{\ii \jj}| < 2a$, otherwise, it is $\rm 0$. $\rm{\mu_{t}}$ is the
motility of an ABP. $\rm \etaT_\ii(t)$ and $\rm \etaR_\ii(t)$ are translational and
rotational noise, respectively. We neglect translational noise in this work to more clearly elucidate the effects of active particle motility and imposed flow on phase separation.
Rotational noise obeys a Gaussian distribution with zero mean
and auto-correlation functions are given by, 
\begin{equation}\label{Rotational_noise_corr}
\rm \langle \etaR_{\ii}(t) \etaR_{\jj}(t') \rangle = \rm 2 \nu_{r} \delta_{\ii \jj}
\delta(t-t')
\end{equation}
where, $\rm \nu_{\rm r}$ is the rotational diffusivity.
$\uu$ and $\rm \bf{\omega}$ represent the
velocity and vorticity fields of the background flow, respectively.

We solve $\rm {2D}$ incompressible Navier-Stokes equation(NSE) to mimic the background flow in which ABPs are moving. We use stream function-vorticity formulation of NSE:~\cite{pandit2017overview} 
\begin{equation}{\label{NSE}}
\rm{ \partial_{t}\bf{\omega}+ \uu .\bf \nabla \bf{\omega}= \rm \nu \bf \nabla^{2}\bf {\omega}-\mu\bf{\omega}+\bf{f_{\omega}}}; \quad
\rm{\bf {\nabla^{2}} \psi = \rm {\omega}};
  \end{equation}
where, ${\rm \psi({\bf{r}}_i,t)}$ and ${\rm \bf {\omega} = \rm \nabla \times \uu(\xx ,t) \equiv \omega \bf \hat{z}}$, are the stream function and vorticity field, respectively, in two dimensions and $\rm {\bf \hat{z}}$ is a unit vector normal to the fluid film. $\rm {\nu}$ is kinematic viscosity and $\rm {\mu}$ Ekman-friction~\cite{gupta2015two,gupta2014elliptical,pandit2017overview}. We take roll-mill form for the external forcing, $\rm \bf{f_{\omega}}$~\cite{gutierrez2019proper,lee2007microfluidic},
\begin{equation}
\rm {\bf{f_{\omega}}} = \rm{{f_0}{\sin(k_f x)\sin(k_{f}y)} \hat{\bf z}}
\end{equation}

where, $\rm {f_{0}}$ is the amplitude of the forcing and $\rm k_f$ is the forcing
wave vector. For large viscosity limit (small Re), and given forcing we will have a steady-state solution for the Navier-Stokes equation.
The exact solution of
the NSE is given by
\begin{equation}\label{eq:Direct_NSE_soln}  
 \rm {{(u_x,u_y)} = \rm \frac{f_0}{4\nu k_f^3}(-sin(k_f {x})cos(k_f {y}), cos(k_f {x})sin(k_f {y})).}
 \end{equation}
 The corresponding vorticity field is
shown in Fig.~\ref{fig:Traj}. 

We use a square box $\rm [0, L]^2$, where box size $\rm L = 2\pi$, with periodic boundary conditions(PBCs) for
background flow and particle dynamics. The square box is uniformly discretized
in $\rm {512^2}$ grid point. In our numerical simulation, we use the
pseudo-spectral method to solve NSE (\ref{NSE}) on these grid points to describe
the dynamics of background flow
\cite{gupta2015two,gupta2014elliptical,pandit2017overview}. Runge-Kutta method
of $\rm{2^{nd}}$ order is used for time marching. We use $\rm k_f = 1$, $\rm \nu = 0.1$,
$\rm \mu = 0$ which produces a four-roll-mill flow, i.e., pairs of clockwise and
anti- clockwise vortices separated by small regions dominated by
strain~\cite{torney2007transport} (see Fig.~\ref{fig:Traj}). This flow has been
studied both experimentally and theoretically for the mixing of passive tracers
~\cite{solomon2003uniform,solomon1988chaotic}. All subsequent analysis for ABPs assumes a steady-state flow (Eq.\ref{eq:Direct_NSE_soln}).

\begin{figure*}[!htb]
\includegraphics[width=1.0\linewidth]{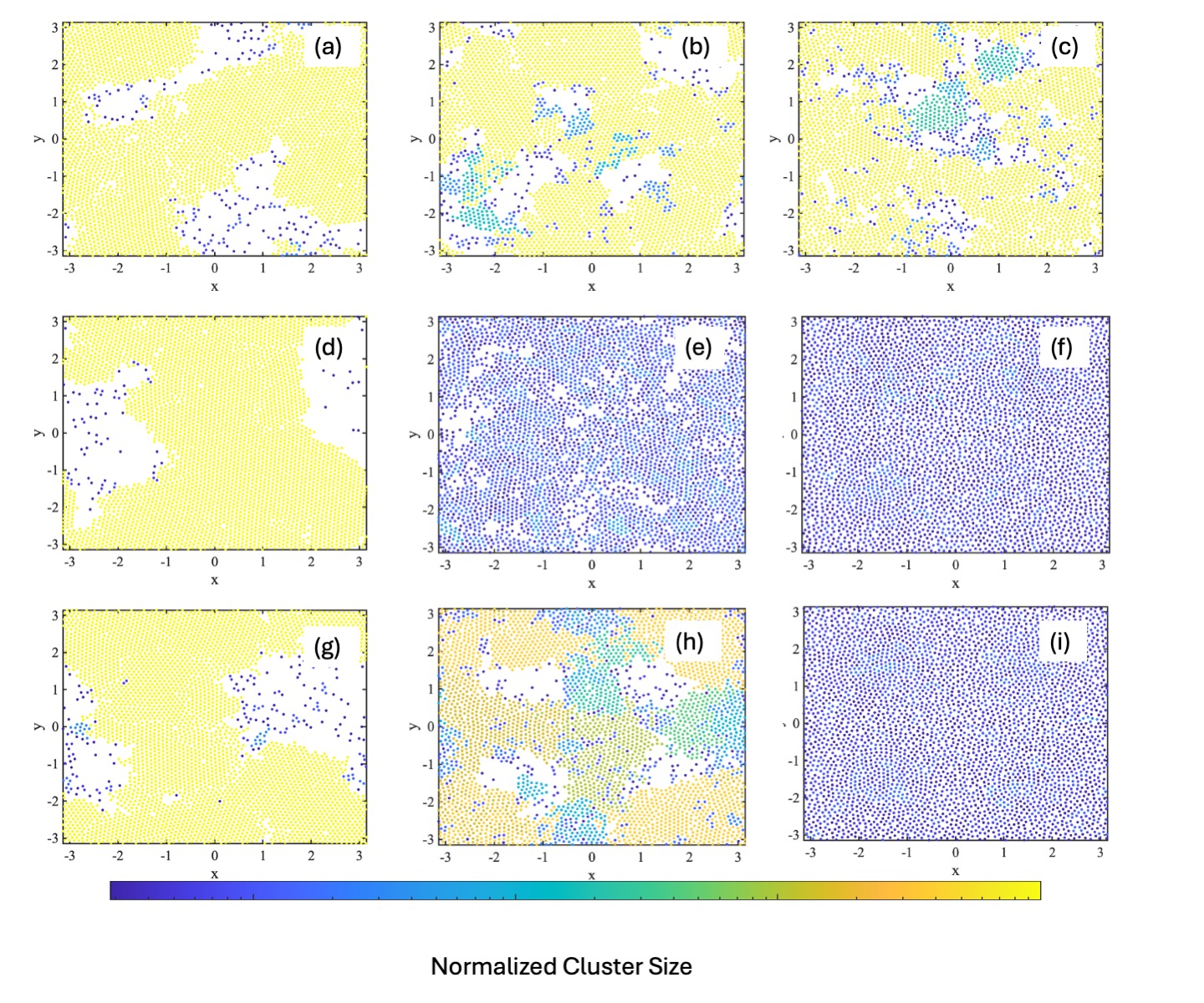}
\caption{\label{fig:snap_plot}\justifying{Representative snapshot of ABPs with $\rm \phi = 0.7$ for three different scaled-speeds {$\rm V $} = 40 (left column), 0.4 (middle column), and 0.004 (right column); and no background flow with $\rm \nu_r = 1.25\times 10^{-5}$ (top row), scaled time $\rm \tau \sim 1$ with background flow (middle row), scaled time $\rm \tau \sim 400$ with background flow (bottom row). The color codes represent the normalized size of the cluster.}}
\end{figure*}

We use bilinear interpolation to determine flow properties at particle
positions. Equations (\ref{position_tem}) and (\ref{theta_tem}) are solved for
$\rm N = 3518$ ABPs, corresponding to a packing fraction $\rm \phi = \rm {0.7}$, to
assess particle dynamics in the background flow using the Euler-Maruyama method
~\cite{malakar2020steady, janzen2022aging}. The packing fraction is defined as
$\rm {\phi = \rm \frac{N \pi \times a^2}{L^2}}$. We have chosen $\rm \phi = 0.7$ because
this $\rm \phi$ value shows MIPS in the absence of background flow~\cite{Fily2012}.
We have kept $\rm \mu_t = 100$ and $\rm k=1$. We are not taking into account the
back-reaction of the ABPs on the flow, and hence no hydrodynamic interaction between them. We use the maximum absolute value of the vorticity $\rm \omega_f = \rm
\frac{f_0}{2\nu k_f^2} =  0.005$ and speed ${\rm{u}_f} =\rm \frac{f_0}{4\nu
k_f^3}= 0.0025$ of the background flow as the characteristic flow vorticity and
speed, respectively. The time scale of the flow is $\rm \sim 1/\omega_f$. Based on the time
\& velocity scales of ABPs and flow, we define dimensionless scaled time $\rm
\tau = \omega_f/\nu_r$ and scaled speed ${\rm V = \vp/\rm{u}_f}$. By considering
the characteristic time and velocity scales of ABPs, we can determine their
persistence length, calculated as $\rm \vp / \nu_r$. This value provides an
estimate of the distance an ABP travels before its direction significantly
changes. All quantitative results presented in this manuscript pertain to three
sets of self-propulsion speeds $\vp$ and three sets of rotational diffusivities
$\rm \nu_r$, yielding scaled times ($\rm \tau = \omega_f/\nu_r$) of $\rm 400.0$
($\rm \gg 1$), $\rm 1$, and $\rm 0.05$ ($\rm \ll 1$), and scaled speeds (${\rm V
= \vp/\rm{u}_f}$) of $\rm 40, 0.4, \& ~ 0.004$. To explore the parameter space
for the phase diagram, we have varied the ${\rm V}$ and $\rm \tau$ across orders
of magnitude. In this study, we fixed all flow parameters and have varied $\rm
\tau$ and ${\rm V}$ by changing $\rm \nu_r$ and $\vp$. For all cases that we consider, we run simulations until the Mean Squared Displacement (MSD) exhibits linear behavior in time.

\section{Results}\label{section3}

We observe three distinct spatial distributions of ABPs across the considered
ranges of $\rm \tau$ and {$\rm V$}. (a) Clustering of ABPs due to MIPS ({$\rm V$} $\rm \gg
1$): In this regime, the ABP velocity significantly exceeds the characteristic
flow velocity, resulting in ABP motility dominating the dynamics and leading to
MIPS. (b) Clustering due to flow topology; flow-induced phase separation (FIPS)
({$\rm V$} $\rm \sim 1$ \& $\rm \tau > 1$), and (c) a homogeneously mixed phase ({$\rm V$}
$\ll 1$). We illustrate these different phases of ABP distribution through a
$\rm \tau-{\rm V}$ phase space (Fig. \ref{fig:comp_phase}). Subsequent subsections
detail qualitative and quantitative comparisons with the corresponding case
without background flow.

\subsection{Phase separation of ABPs}\label{Phase_sep}
In the absence of background flow, a dense suspension ($\rm \phi \geq 0.4$) of
interacting ABPs exhibits the well-known phenomenon of
MIPS~\cite{Fily2012,redner2013structure}. The top row of
Fig.~\ref{fig:snap_plot} shows snapshots of particle positions for three
different self-propulsion velocities $\rm \vp$: $0.1$ (panel (a)), $1 \times 10^{-3}$ (panel (b)), and $1 \times 10^{-5}$
(panel (c)). In all cases, $\rm \phi=0.7$ and $\rm \nu_r$ is fixed at $1.25 \times 10^{-5}$. 

\begin{figure*} 
\includegraphics[width=0.49\linewidth]{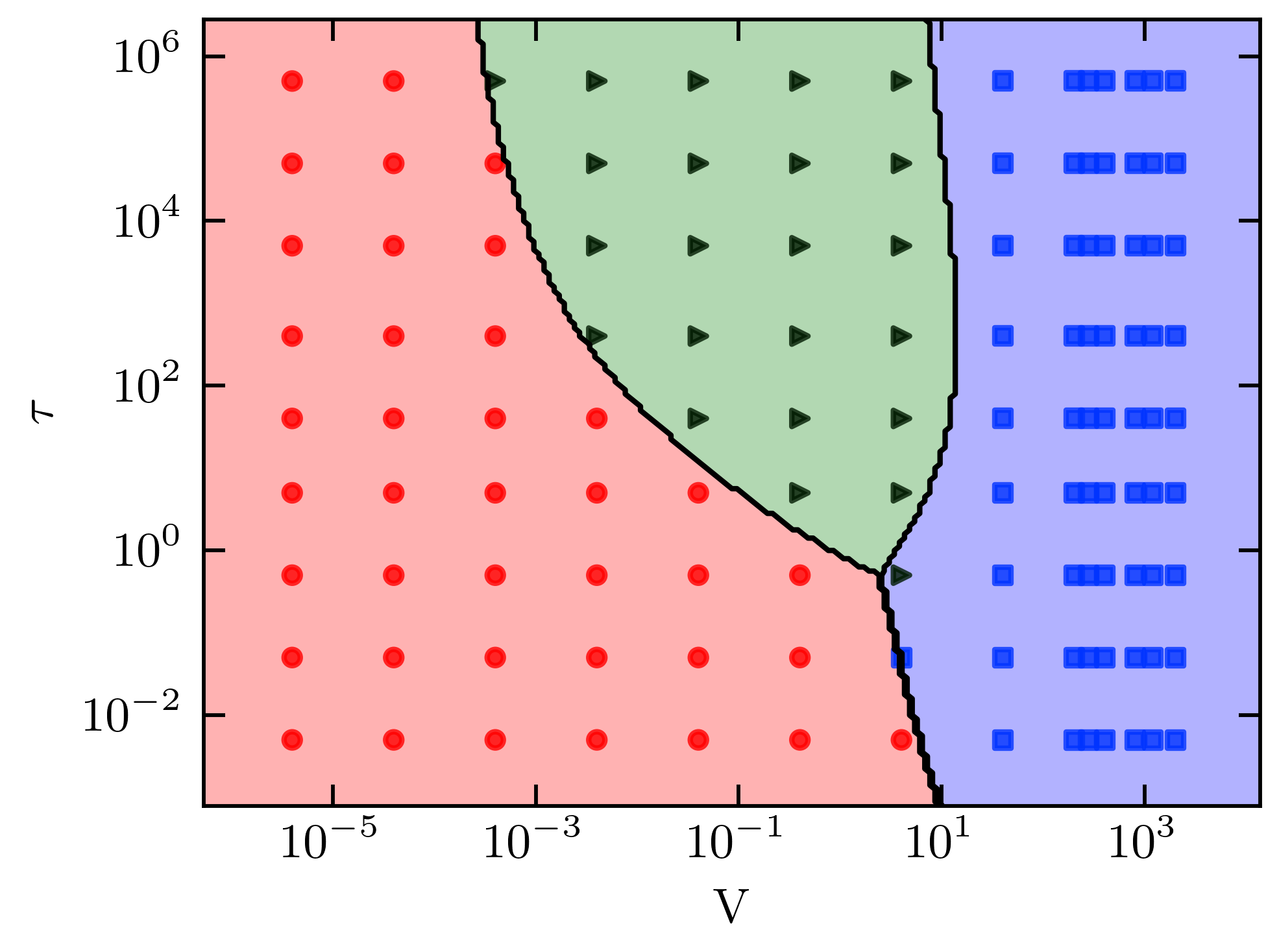} 
\put(-60,160){ { \textcolor{black}{\normalsize \bf (a)} } }
\put(-205,85){ { \textcolor{black}{\normalsize \bf Homogeneous} } }
\put(-130,150){ { \textcolor{black}{\normalsize \bf FIPS} } }
\put(-50,95){ { \textcolor{black}{\normalsize \bf MIPS} } }
\includegraphics[width=0.49\linewidth]{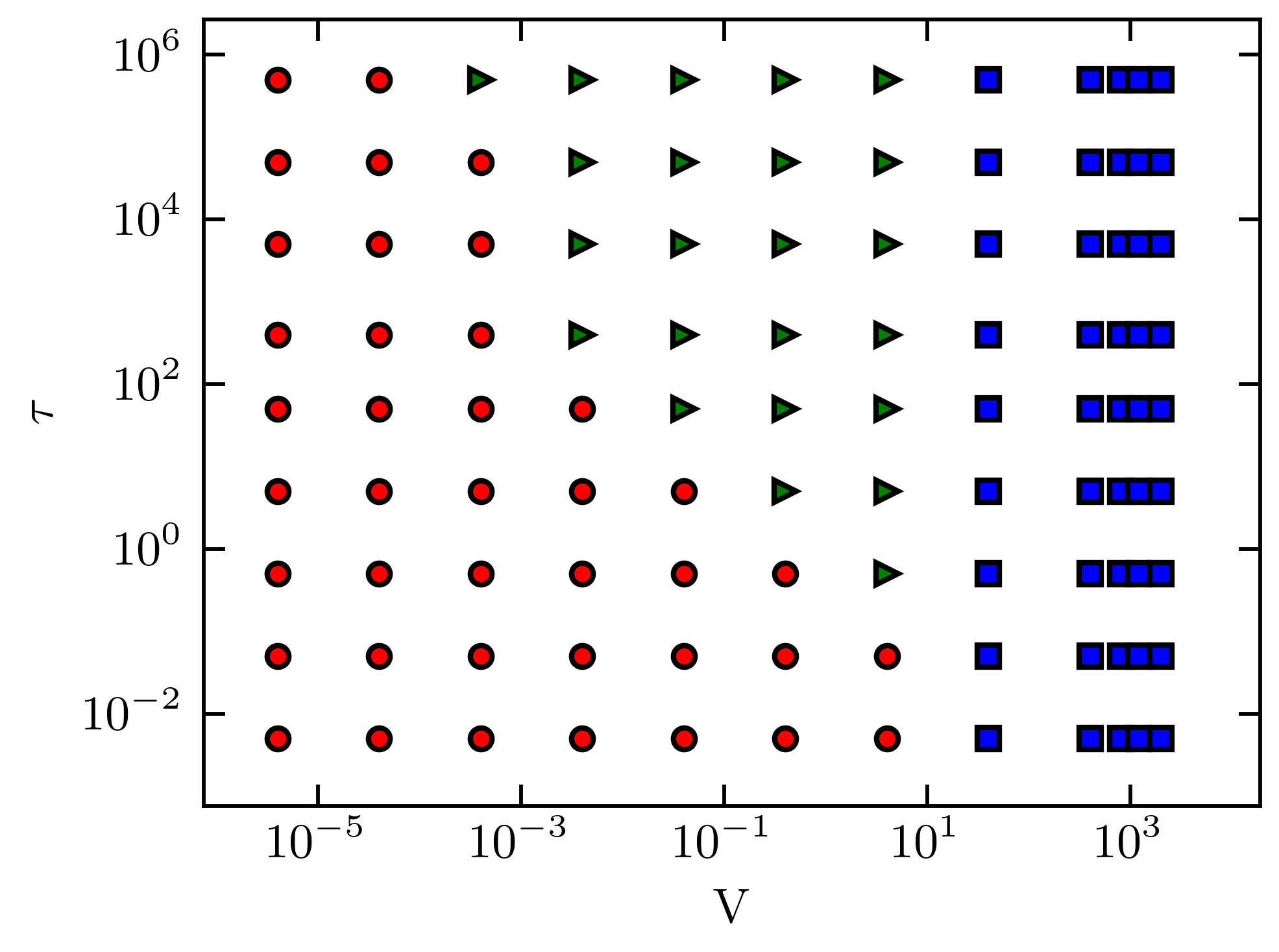}
\put(-60,160){ { \textcolor{black}{\normalsize \bf (b)} } }
\caption{\label{fig:comp_phase}\justifying{(a) Phase diagram of ABPs spatial distribution in $\rm \tau$ - $\rm{V}$ plane for $\rm \phi = 0.7$. Three different phases --- homogeneous phase (red region), flow-induced phase separation (FIPS) (green region), and motility-induced phase separation (MIPS) (blue region) --- are observed. The dots represent actual simulations and the shaded background corresponds to best-fit predictions for regime boundaries using support vector machines (SVM). (b) $\OP$ in the $\rm \tau$ - $\rm{V}$ plane. Red circles correspond to homogeneous phase ($\OP = 1$), green triangles corresponds to FIPS phase ($\OP = 2$) and blue square 
correspond to MIPS phase ($\OP = 0$)}}
\end{figure*}
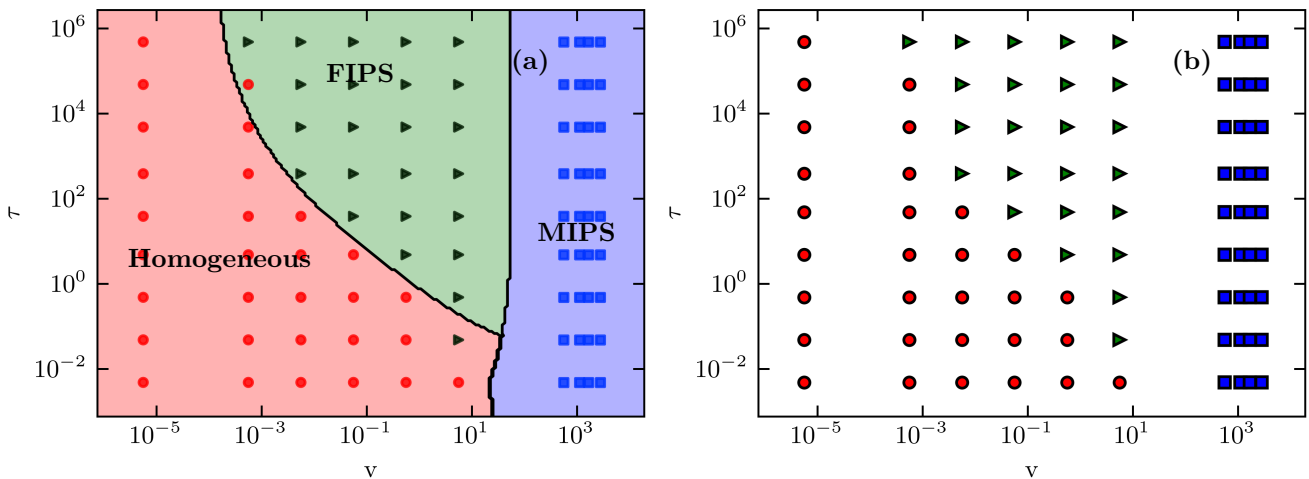

When a background (roll-mill) flow is introduced to these suspensions, we
observe that MIPS persists for cases with large scaled speed (i.e., $\rm V
\gg 1$), regardless of the scaled time $\rm \tau$. This is shown in the panels (d)
and (g) of Fig.~\ref{fig:snap_plot} for $\rm \tau = 1$ and $\rm \tau = 400$,
respectively. In this regime, the background flow acts as a small perturbation
to the dynamics of ABPs, evident from the red trajectory shown in
Fig.~\ref{fig:Traj}, and MIPS is still dominant. When we decrease the scaled
speed $\rm V$ by decreasing the self-propelled velocity $\vp$ and making it
comparable to the fluid velocity scale ($\rm V ~\sim 1$), we observe that the
clustering is dependent on the scaled time $\rm \tau$. When scaled time $\rm \tau \le 1$
we observe a weak clustering phase as shown in Fig.~\ref{fig:snap_plot}(e). For
$\rm \tau > 1$, we observe a phase-separated state different from MIPS. The flow
topology drives this clustering and all the ABPs try to cluster themselves in
the strain-dominated region, i.e., the region with Okubo-Weiss parameter
$\rm \Lambda < 0$ as shown in Fig.~\ref{fig:snap_plot}(h). One can observe in
Fig.~\ref{fig:Traj} that the blue trajectory is slowly drifting away from the
vortex core and towards the strain-dominated region. Okubo-Weiss parameter is
defined as $\rm \Lambda = \rm \frac{\omega^2 -\sigma^2}{8}$
~\cite{gupta2015two,perlekar2011persistence}, $\rm \Lambda > 0$ and $\rm < 0$ represent
vortical and strain-dominated regions, respectively, where $\rm \sigma^2 \equiv
\sum_{ij} \sigma_{ij}\sigma_{ij}$ (strain rate), and
$\rm \sigma_{ij}\equiv\partial_i u_j + \partial_j u_i$; $\rm i,j$ are cartesian-indices.
It is important to note that we observe this type of clustering only when the speed of ABPs and background flow are comparable and $\rm \tau \gg 1$. Since,
$\rm \tau \gg 1$, the persistence length for ABPs is large compared to the vortical
structures (half of the box size in this case) of the flow, and once these
particles reach the strain-dominated region, which are straight paths for this
flow, tend to stay in this region. If the ABPs are in the vortical region, due
to the long persistence length of the ABPs, they exit the rotating vortical
regions and are pushed to the strain-dominated region. This clustering is due to
the flow topology and the finite size of the ABPs, so we define this clustering
behavior as flow-induced phase separation (FIPS). When we further decrease the
scaled velocity ${\rm V} \ll 1$, the mixing due to the roll-mill flow starts to
dominate, and we see the system in the homogeneous phase, for both $\rm \tau$ values
as shown in Figs. ~\ref{fig:snap_plot}(f) and ~\ref{fig:snap_plot}(i). 

We have observed that the system's state --- MIPS, FIPS, or homogeneous phase
--- depends on the scaled velocity {$\rm V$} and scaled time $\rm \tau$. To elucidate
this further, we conducted a parameter scan in $\rm \tau$-{$\rm V$} space
(Fig.~\ref{fig:comp_phase}). For small $\rm \tau$, systematically increasing {$\rm V$}
transitions the system from a homogeneous state (red region) to a MIPS state
(blue region). Upon exceeding $\rm \tau = 1$, a new phase-separated state, FIPS
(green region), emerges. The {$\rm V$} range exhibiting FIPS expands with
increasing $\rm \tau$. At extremely large {$\rm V$}, only MIPS is observed. Within the
FIPS regime, most ABPs reside in strain-dominated regions where $\rm \Lambda < 0$.
We can characterize $\rm \bar {\Lambda} \equiv 1/N \sum_{i} \langle \Lambda_i
\rangle_t$ for ABPs in the steady state, where $\rm \langle \rangle_t$ represents
time average. Based on the sign of $\rm \bar \Lambda$ we can determine whether a
regime in FIPS ($\rm \bar \Lambda <0$) or MIPS/homogeneous ($\rm \bar \Lambda > 0$)
(Appendix: Fig.~\ref{fig:Lambda}(a)).

In MIPS dominated regime, we do not expect any correlation between the flow
velocity and the ABP velocity. It is given by  
\begin{equation}\label{correlation_uv}
\rm \uv \equiv \frac 1 N
\sum_{i} \left \langle \frac{{\bf u}({\bf r}_i,t) \cdot {\bf v}_i} {|{\bf u}({\bf
r}_i,t)| |{\bf v}_i)|} \right \rangle_t
\end{equation}
So if we estimate the equation \ref{correlation_uv}, it will be close to zero in the MIPS regime
and close to 1 for the homogeneous regime. $\uv$ will maintain an intermediate
value for FIPS regime (Appendix: Fig.~\ref{fig:Lambda}(b)). Hence, by combining the above
information, we can define the following order parameter: $\OP$ and is given by,
\begin{equation}
\OP = \rm \Theta(-\bar
\Lambda) + \Theta(\uv-\epsilon)
\end{equation}
where $\rm \epsilon$ is a threshold value that we
have chosen to be $\rm 0.02$ to differentiate the $\uv$ value from MIPS to the rest
of the regimes. Fig.~\ref{fig:comp_phase}(b) shows $\OP$ in $\rm \tau$ - $\rm V$
plane. $\OP = 0$ (blue square) corresponds to the MIPS regime, as both the terms
in the $\OP$ will not contribute. $\OP = 1$ (red circles) corresponds to the
homogeneous regime, as only the second terms in the $\OP$ will contribute. $\OP
= \rm 2$ (green triangles) corresponds to the FIPS regime, as both the terms in the
$\OP$ will contribute. In Fig.~\ref{fig:comp_phase}(b) we can observe different
values of order parameter $\OP$ for different phases similar to
Fig.~\ref{fig:comp_phase}(a) where different phases are marked with different
colors based on the distribution of particles. 
In Fig.~\ref{fig:comp_phase}(a), we manually screened steady-state snapshots across phase space to clearly identify the three distinct regions.

\subsection{ Mean Square Displacement (MSD)}

For interacting ABPs, the mean-square displacement (MSD) is given by~\cite{Fily2012}
\begin{equation}\label{eq:MSDful}
\rm \langle[\Delta \boldsymbol{r}(t)]^2\rangle = \rm 4 D_{e} \left[t + \frac{1}{\nu_r}(e^{-\nu_r t}-1) \right],
\end{equation}
where $\rm D_e = \rm {v_0}^2/2\nu_r$ is the effective diffusivity and $\rm v_0$ is
effective drift velocity; for non-interacting ABPs $\rm v_0 = v_p$.

\begin{figure*}[!htb]
\includegraphics[width=0.33\linewidth]{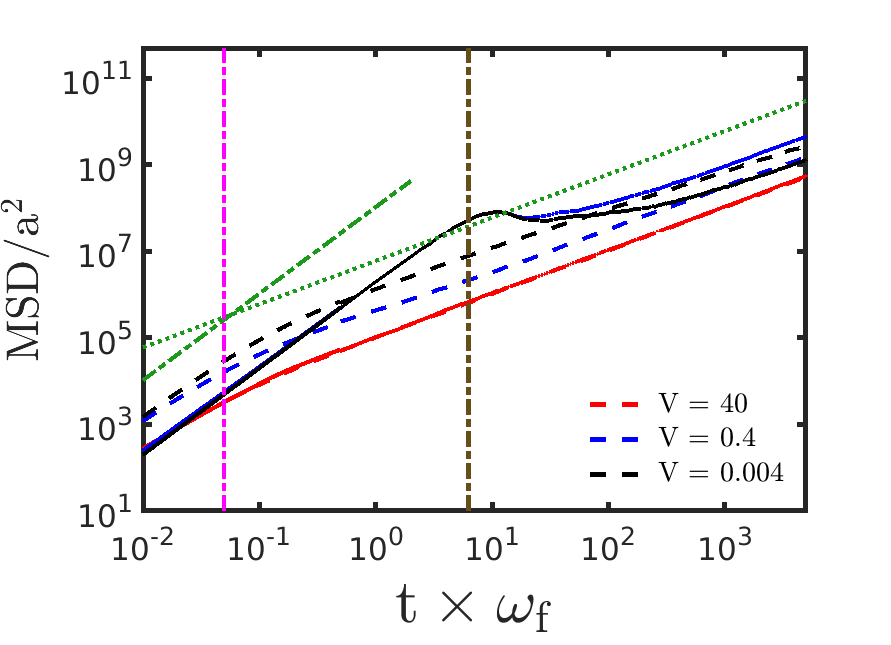}
\put(-135,100){ { \textcolor{black}{\normalsize \bf (a)} } }
\includegraphics[width=0.33\linewidth]{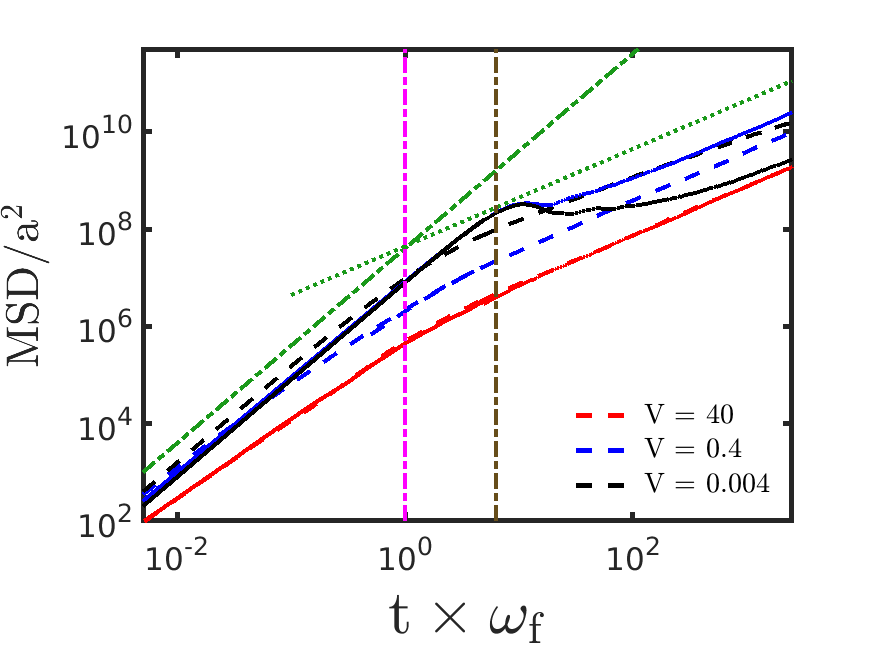}
\put(-135,100){ { \textcolor{black}{\normalsize \bf (b)} } }
\includegraphics[width=0.33\linewidth]{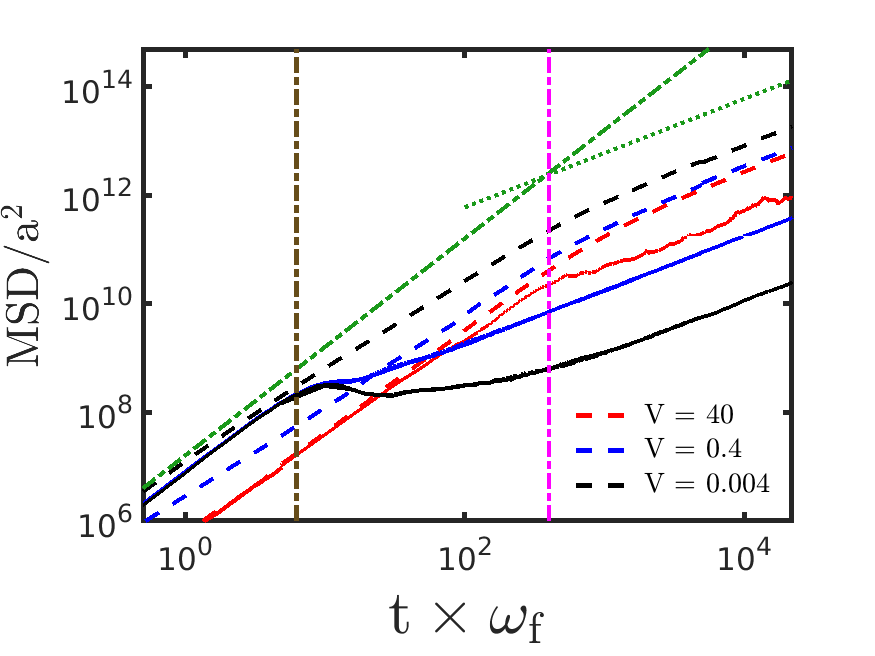}
\put(-135,100){ { \textcolor{black}{\normalsize \bf (c)} } }
\caption{\label{fig:msd}\justifying{Normalized Mean-square displacement (MSD) of ABPs
versus normalized time with $\rm \phi =0.7$ and for scaled-time (a) $\rm \tau = 0.05$, (b) $\rm \tau
= 1$ \& (c) $\rm \tau = 400$. Solid (dashed) lines indicate MSD with (without)
background flow for three different scaled-speeds $\rm V = $ 40 (red
curve), 0.4 (blue curve), \& 0.004 (black curve). Length is non-dimensionalized by the size of the ABPs, and time by the fluid timescale, $\omega_f^{-1}$. The green dashed-dotted (dotted)
line shows the slope 2 (1) on a log scale, guides to the eyes for a ballistic (diffusive)
 regime. Their intersection shows the transition time $\Ttrans$ from
ballistic to diffusive regime. The dashed-dotted magenta and brown vertical lines represent the rotational diffusion time scale ($\rm 1/\nu_r$) in scaled time unit at $\tau$ and a fluid time scale ($\rm 2\pi/\omega_f$) in scaled time unit at $2 \pi$, respectively}. To bring MSDs for different $\rm V$ on the same y-scale, we rescale the MSDs by $\rm 1/{v}_p^2$ for MIPS cases and by $\rm 1/{u_f}^2$ for FIPS and homogeneous cases.}
\end{figure*}

\begin{figure*}[!htb]
  \includegraphics[width=0.50\linewidth]{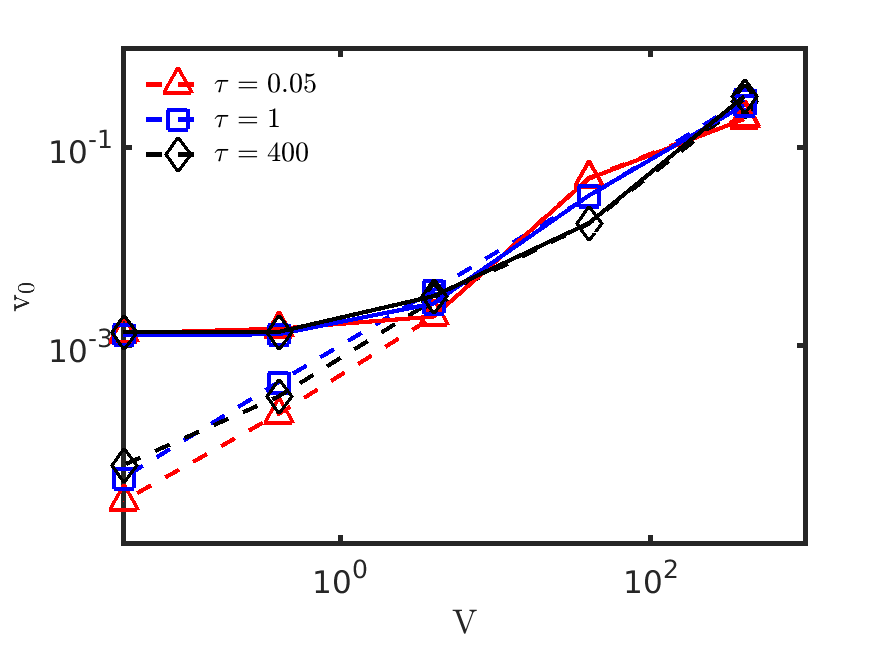}
  \put(-50,50){ { \textcolor{black}{\large \bf (a)} } }
  \includegraphics[width=0.50\linewidth]{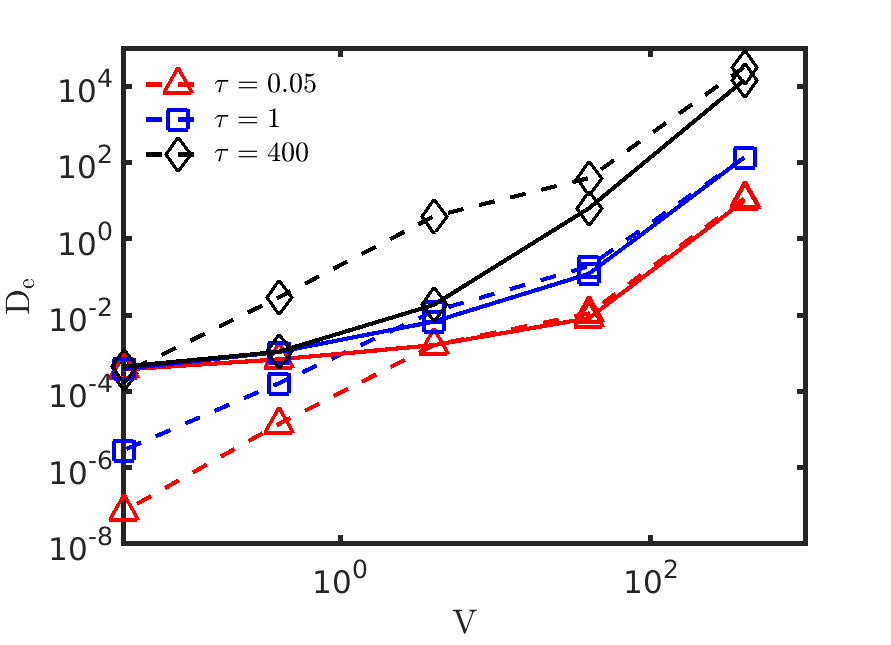}
  \put(-50,50){ { \textcolor{black}{\large \bf (b)} } }
\caption{\label{fig:diffusioon_drift}\justifying{(a) Drift velocity ($\rm {v_0}$) and (b) Diffusivity ($\rm D_e$) of ABPs versus scaled-speed $\rm V$ with $\rm \phi = 0.7$ for three different scaled times $\rm \tau \sim 0.05$ (red), $\sim 1$ (blue), $\sim 400$ (black). Solid (dashed) lines indicate with (without) background flow.}}
\end{figure*}

In Fig.~\ref{fig:msd} we show the MSD for ABPs, where the solid (dashed) lines
are for the case with (without) background flow. As expected, for $\rm {t \ll \nu_{r}^{-1}}$, there is ballistic behavior with $\rm { \big
\langle[\delta r(t)]^{2} \big \rangle \sim v_{0}^{2} t^{2}}$, whereas at later
time, the diffusive behavior with $\rm {\big \langle[\delta
\boldsymbol{r}(t)]^{2} \big \rangle \sim 4 D_{e} t}$ is observed. 
The time of cross-over from ballistic to diffusive regime is defined as transition time $\Ttrans$.
We show the MSD for three scaled-times $\rm \tau \ll 1$
(Fig.~\ref{fig:msd}(a)), $\rm \tau \sim 1$ (Fig.~\ref{fig:msd}(b)) \& $\rm \tau \gg 1$
(Fig.~\ref{fig:msd}(c)). Each panel shows the plots for three scaled-speeds
${\rm V} \gg 1$ (red), $\simeq 1$ (blue), $\ll 1 $(black). We observe that for
the case without background flow, the transition time $\rm \Ttrans \simeq
\nu_r^{-1}$, marked by vertical magenta lines, whereas for the case with
background flow, the $\Ttrans$ is $\rm V$ dependent. For large $\rm V$,
$\Ttrans$ is independent of background flow (red-solid curve) as MIPS drives the
dynamics. For $\rm V \ll 1$ (Fig.~\ref{fig:msd} black-solid curve), since the
major contribution to the velocity of the ABPs is the background flow, $\Ttrans$
is dependent on the fluid time scale, marked by the brown vertical lines. 
$\Ttrans$ for the FIPS case (blue-solid curve) follows a similar trend as $\rm V \ll 1$ case,
with a slight delay. 
An important point to note is that the
MSDs for FIPS and mixed-phase display a bump before going to the diffusive
regime. This bump is associated with the fact that the ABPs in this regime
follow the streamlines and move in quasi-closed orbits. The bump for the mixed
phase is stronger as they follow the closed orbits for a relatively longer time,
attributed to their low motility Fig.~\ref{fig:Traj}. A similar bump
is observed in MSD for the non-interacting ABPs~\cite{caprini2020diffusion},
attributed to the trapping effect due to the vortices of the roll-mill flow.

\begin{figure*}[!htb]
  \includegraphics[width=0.49\linewidth]{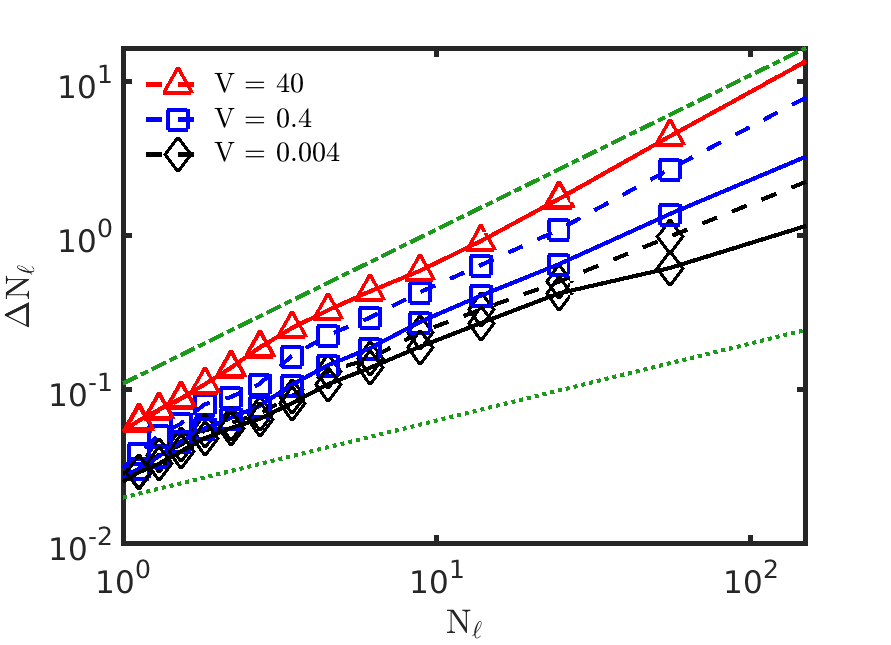}
  \put(-50,50){ { \textcolor{black}{\large \bf (a)} } }
  \includegraphics[width=0.49\linewidth]{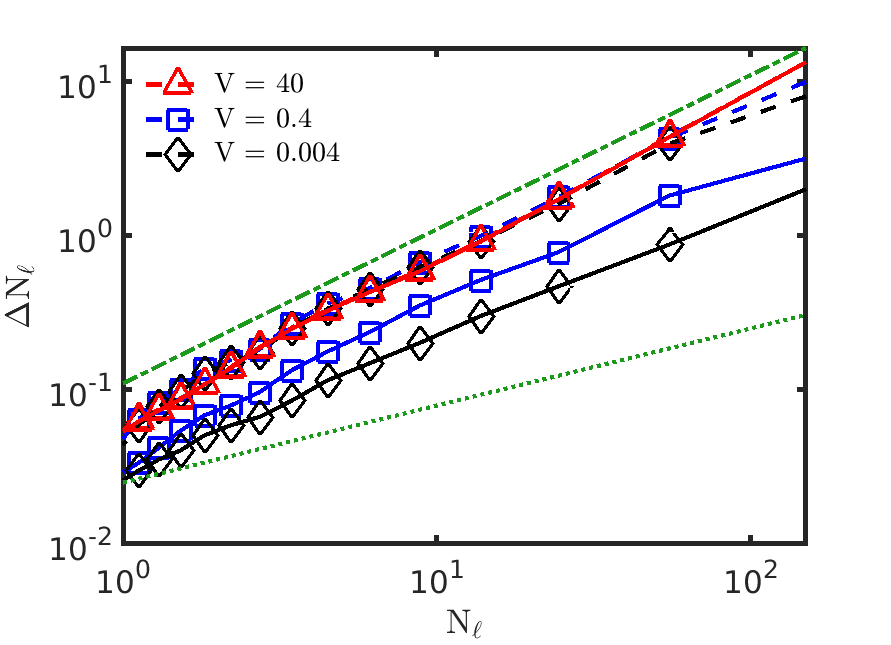}
  \put(-50,50){ { \textcolor{black}{\large \bf (b)} } } 
\caption{\label{fig:stad_dev}\justifying{Standard deviation of fluctuations $\rm{\Delta N_\ell}$ versus average number $\rm{N_\ell}$ of ABPs in subsystem of size $\ell=\rm \sqrt{\pi a^2 N_\ell/\phi}$ for scaled-time (a) $\rm \tau \sim 1$ \& (b) $\tau \sim 400$. Solid (dashed) lines are for ABPs with (without) background flow for three different scaled speeds $\rm{V} = $ 40 (red curve), 0.4 (blue curve), \& 0.004 (black curve). The green dashed-dotted (dotted) line denotes $\rm {\Delta N_\ell = N_\ell}$ ($\rm {\Delta N_\ell = N_\ell^{\frac{1}{2}}}$}).}
\end{figure*}

{\it Drift velocity.}- Using
eq.~\ref{eq:MSDful}, in the limit $\rm t \ll \Ttrans$ we extract the drift velocity
$\rm \vd$ from MSD (Figure ~\ref{fig:diffusioon_drift}(a)). In
Fig.~\ref{fig:diffusioon_drift}(a), we show that plot of drift velocity $\rm \vd$ vs
scaled speed ${\rm V}$. The drift velocities for the case without background
flow (dashed curves) go together for different values of $\rm \tau$, as there is no
perturbation from the background flow $\rm \Rightarrow \vd \sim {\rm v}_p$, as can
be seen in Fig.~\ref{fig:diffusioon_drift}(a), the dashed straight lines with
slope 1, for $\rm \tau \sim$ 0.05, 1 \& 400. Once we introduce the background flow
(solid curves), we see that for $\rm V < 1$ the fluid velocity dominates, and
the drift velocity of ABPs is comparable to fluid speed $\rm u_f$. This
behaviour is independent of $\rm \tau$, and $\rm \vd$ remains comparable to $\rm u_f$
till ${\rm V}$ approaches unity. For large ${\rm V}$ the drift velocity again
becomes comparable to the $\rm v_p$.

{\it Diffusivity.}- Similarly, using eq.~\ref{eq:MSDful}, in the limit $\rm t \gg
\Ttrans$, we extract diffusivity $\rm D_e$ from MSD. In
Fig.~\ref{fig:diffusioon_drift}(b) we show the plot of diffusivity $\rm D_e$ vs scaled speed ${\rm V}$. The diffusivity for the case without background flow (dashed
curves) follows $\rm \vd^2/2\nu_r$, maintains a slope `$\rm 2$' and a gap of the ratio
of the time scales, i.e., $\rm \sim 400$ between black and blue and $\rm \sim 20$
between blue and red curves. 
In the presence of background flow, when ${\rm V} \ll 1$, the diffusivity is
completely controlled by the background flow and it becomes independent of
$\rm \tau$. As ${\rm V}$ starts to approach unity, we observe that the diffusivity
curves start to bifurcate for different $\rm \tau$ values and approach the value
corresponding to the without background flow. Finally, curves merge with their
counterparts in the absence of background flow for ${\rm V} \gg 1$.  

\begin{figure*}
  \includegraphics[width=0.49\linewidth]{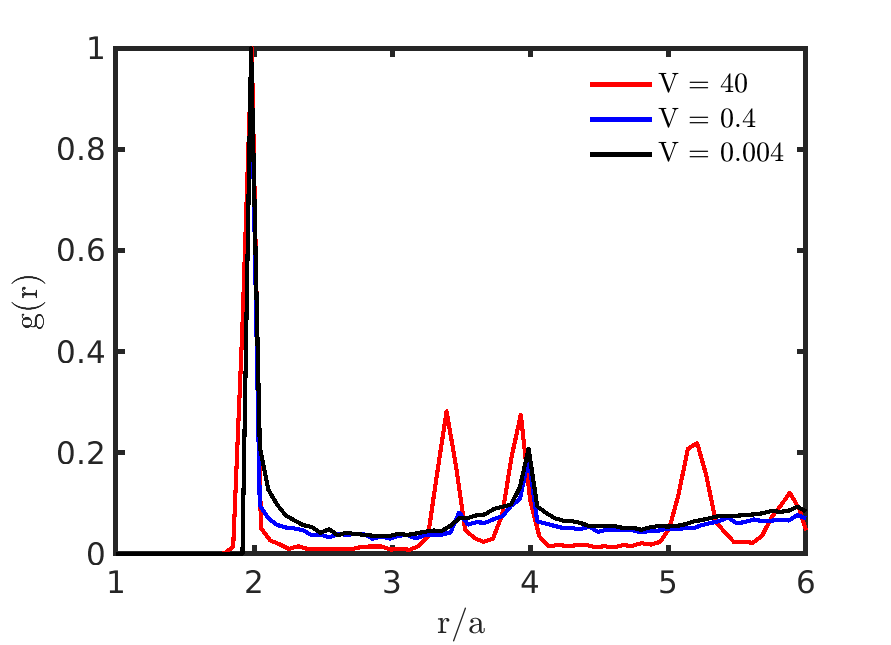}
  \put(-60,115){ { \textcolor{black}{\large \bf (a)} } }
  \includegraphics[width=0.49\linewidth]{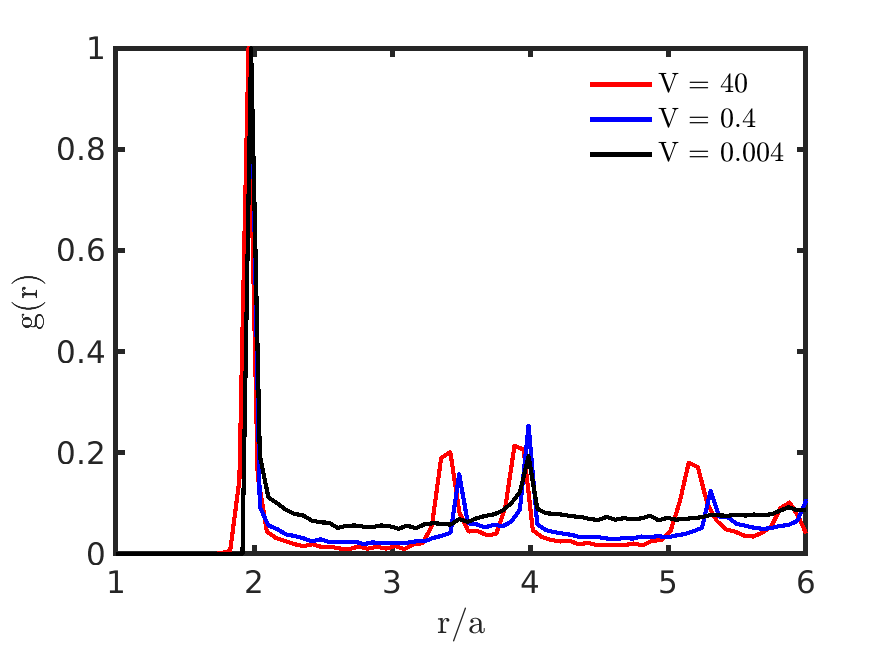}
  \put(-60,115){ { \textcolor{black}{\large \bf (b)} } }
\caption{\label{fig:rad_plot}\justifying{Radial distribution function (RDF) $\rm{g(r)}$ versus $\rm {r/a}$ of ABPs for fixed $\rm{\phi =0.7}$, for three different scaled speeds $\rm V = $ 40 (red curve), 0.4 (blue curve), \& 0.004 (black curve) with scaled time (a) $\rm \tau \sim 1$ and (b)$\rm \tau \sim 400$.}}
\end{figure*}

\subsection{Giant number fluctuation}

In phase-separating active systems, we anticipate observing giant number fluctuations, characterized by the relationship $\rm {\Delta N_\ell = \rm N_\ell^\alpha}$, where $\rm {N_\ell}$ and $\rm {\Delta N_\ell}$ represent the average and variance, respectively, of the number of Active Brownian Particles (ABPs) within a subdomain of size $\rm \ell^2$. For equilibrium systems with a homogeneous uniform distribution, the exponent $\alpha$ is 0.5. However, as the system deviates from equilibrium and exhibits a non-uniform distribution, $\alpha$ can approach $1$ in two dimensions~\cite{Fily2012,henkes2011active,Narayan2007,kuroda2023anomalous}.

Specifically, in Motility-Induced Phase Separation (MIPS) with ABPs, $\alpha$ has been reported to approach 1 with increasing Péclet number ($\rm Pe$) for packing fractions $\phi>\phi_c$, and with increasing $\phi$ for large $\rm Pe$~\cite{Fily2012,henkes2011active}. For two-dimensional nematic systems, the giant number fluctuation exponent transitions from $\alpha = 0.5$ to $\alpha = 1$ as the packing fraction $\phi$ increases from 0.35 to 0.66, indicating a shift from a uniform~\cite{Narayan2007,ramaswamy2003active}. Inertial Active Brownian Particles (iABPs) exhibit $\alpha = 0.5$ for $\rm Pe < 10$, corresponding to the homogeneous region, and $\alpha = 0.9$ for the phase-separated region when $\rm Pe \ge 100$~\cite{kuroda2023anomalous}. These giant number fluctuations serve as a signature of symmetry breaking induced by the active velocity term in the system.

 We find that, in the
absence of background flow, for all values of {\rm V} and $\rm \tau$, the slope $\rm \alpha$
approaches 1 (dashed curves), indicative of a clear phase-separated state
(Fig~\ref{fig:stad_dev}). In the phase-separated MIPS region for scaled velocity
$\rm V \gg 1$ we observe $\rm \alpha \sim 1$ (red-solid curve), similar to the
case without background flow (red dashed curves). In the FIPS region (the green
region in Fig.~\ref{fig:comp_phase}), where $\rm V \sim 1$ \& $ \rm \tau \gg 1$ we
find $\rm 0.5< \alpha < 1$ (solid-blue curve in Fig~\ref{fig:stad_dev}). Here, ABPs
form clusters in strain-dominated regions, but the clustering is less pronounced
than in the MIPS regime, leading to $\rm \alpha$ values between $\rm 0.5$ and $\rm 1$. In the
mixed-phase (red region in Fig.~\ref{fig:comp_phase}) where $\rm V \ll 1$,
$\rm \alpha$ is approximately $\rm 0.5$, here ABPs are following the background
flow and the exponent corresponds to an equilibrium state.  

\begin{figure*}
\includegraphics[width=0.49\linewidth]{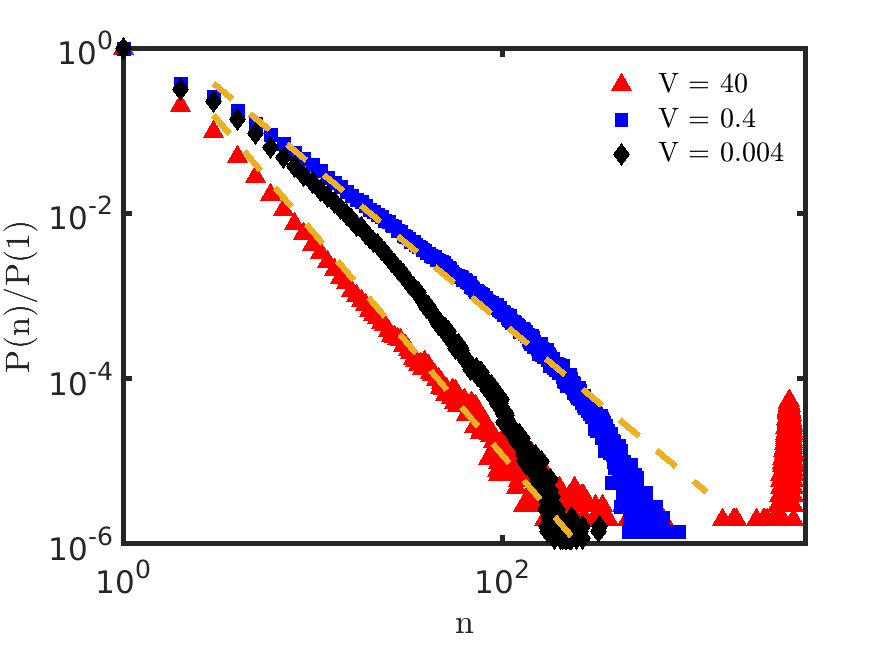}
  \put(-60,115){ { \textcolor{black}{\large \bf (a)} } }
\includegraphics[width=0.49\linewidth]{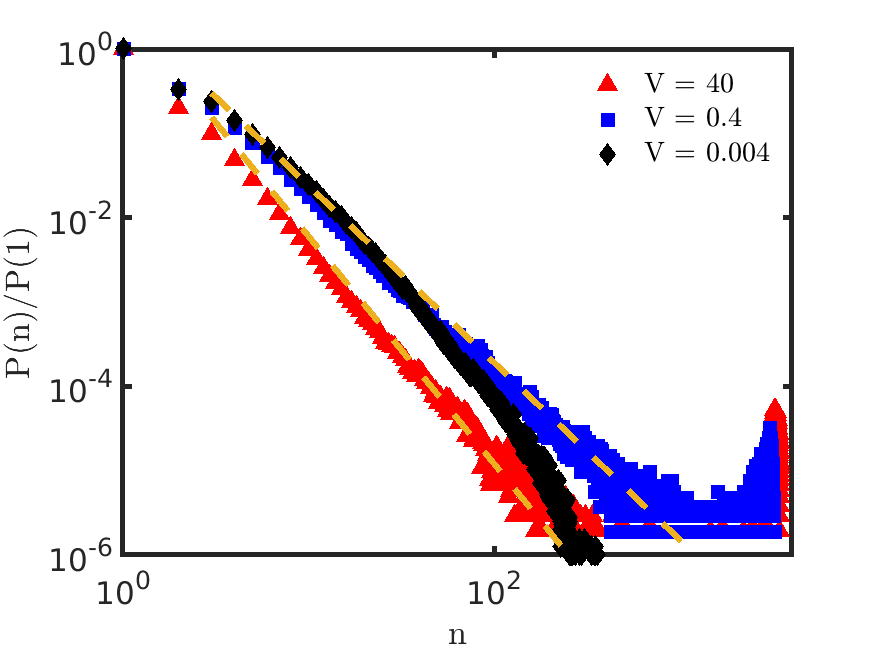}
  \put(-60,115){ { \textcolor{black}{\large \bf (b)} } }
\caption{\label{fig:csd}\justifying{Cluster size distribution (CSD) $\rm {\rm P(n)/P(1)}$ versus $\rm n$ where $\rm n$ is the cluster size for ABPs for fixed $\rm \phi = 0.7$, for three different scaled speeds $\rm{V} = $ 40 (red triangles), 0.4 (blue squares), \& 0.004 (black diamonds) with scaled time (a) $\rm \tau \sim 1$ and (b)$\rm \tau \sim 400$. Two yellow dashed lines guide to the eye with the slopes $-2.7$ \& $-1.9$ respectively in (a), and slopes $-2.7$ \& $-2.1$ respectively in (b).}}
\end{figure*}

\subsection{Radial Distribution Function}

To understand the packing behavior of ABPs in different regimes of the phase
space (Fig.~\ref{fig:comp_phase}), we calculate the radial distribution function
(RDF) as shown in Fig.~\ref{fig:rad_plot}. In the MIPS region, where $\rm V \gg
1$, ABPs within large clusters exhibit close packing with local hexagonal order,
as evident in the first and second peaks at $\rm 2 r/a$, $\rm 2\sqrt3 r/a$,
respectively, of the RDF (Fig.~\ref{fig:rad_plot}(a) and (b), red solid
lines)~\cite{singh2022effective}. In the FIPS region, characterized by $\rm V
\sim 1$ and $\rm \tau \gg 1$, the RDF displays peaks at $\rm 2 r/a$ and $\rm 4
r/a$, with a smaller peak at $\rm 2\sqrt3 r/a$ (Fig.~\ref{fig:rad_plot}(b), blue
curve). This intermediate peak suggests transient hexagonal packing due to the
clustering in the strain-dominate regions of the flow. In contrast, the
homogeneously mixed region, defined by $\rm V \ll 1$ (Fig.~\ref{fig:rad_plot},
black curve) or $\rm V \sim 1$ and $\rm \tau \le 1$ (Fig.~\ref{fig:rad_plot}(a),
blue curve), exhibits RDF peaks at $\rm 2 r/a$ and $\rm 4 r/a$, indicative of a
lack of ordered packing among the ABPs.

\subsection{Cluster Size Distribution}

After understanding packing, we focused on the size of the cluster and its
distribution in a statistically steady state. In Fig.~\ref{fig:csd}, we have
plotted the normalized cluster size distribution (CSD) $\rm P(n)/P(1)$ vs cluster
size $\rm n$, where $\rm P(n)$ is the probability of finding a cluster of size `$\rm n$'
ABPs. Irrespective of the parameter space, normalized CSD fits a power law with
exponential cut-off is given by ~\cite{dolai2018phase},
\begin{equation}
\rm P(n)/P(1) \simeq 1/n^\beta
\exp({-n/n_0})
\end{equation}
where values of exponent $\beta$ are reported in the caption of Fig.~\ref{fig:csd}. In
the MIPS region, where ${\rm V} \gg 1$ (red curve), we observe a bimodal
distribution of cluster-size $\rm n$ (Fig.~\ref{fig:csd}(a-b)), which is similar to
the observation for standard MIPS~\cite{sanoria2022percolation,dolai2018phase}. The bimodal
distribution is a signature of the clear phase separation, and it is also
observed for the FIPS region (blue curve, Fig.~\ref{fig:csd}(b)). For a homogeneous distribution (${\rm V}
\ll 1$ (black curve) and $\rm V \sim 1$
\& $\rm \tau \sim 1$ (blue curve, Fig.~\ref{fig:csd}(a))), CSD shows a unimodal power-law decay. 
The slow decay for $\rm V \sim 1$
\& $\rm \tau \sim 1$ (shown by the dashed yellow line with $\rm \beta = 1.9$, a guide to
the eye), can be attributed to the formation of moderately large clusters, in
comparison to the case with $\rm V \ll 1$. Additionally, for $\rm \tau \sim 1$
(Fig.~\ref{fig:csd}(a)), CSD shows a non-monotonous power-law exponent $\rm \beta$
as a function of $\rm V$. 

\section{Conclusion}\label{section4} 

We investigated the dynamics of interacting ABPs with and without background
flow in two dimensions. We have used a four-roll mill flow, as a background
flow, where the topology remains fixed in space with distinct vortical and strain-dominated regions. While previous studies on non-interacting self-propelled particles in this flow
revealed strong clustering only for anisotropic particles \cite{torney2007transport},
our results demonstrate that isotropic ABPs with steric repulsion exhibit
clustering under certain parameter regimes (Fig.~\ref{fig:comp_phase}). We
identify three distinct phases based on particle spatial distribution:
homogeneous, flow-induced phase separation (FIPS), and motility-induced phase
separation (MIPS). The phase diagram in Fig.~\ref{fig:comp_phase} reveals
precise control over these phases through the scaled speed ($\rm V$) and
scaled time ($\rm \tau$). A novel phase, FIPS, emerges for $\rm \tau > 1$
and $\rm V \sim 1$, characterized by clustering in the strain-dominated
region, distinct from MIPS. This FIPS phase occupies a narrow region in the
parameter space, as illustrated in Fig.~\ref{fig:comp_phase}.

In the MIPS phase, ABPs exhibit no correlation with flow characteristics.
Conversely, in the FIPS phase, they respond to flow topology and correlate with
flow velocity. The homogeneous phase displays a strong correlation with flow
velocity but no discernible correlation with flow topology. Based on these
observations, we introduced an order parameter that distinguishes between the
three phases by adopting distinct values in each regime.

To quantitatively differentiate the phases, we analyzed drift velocity,
diffusivity, mean squared displacement (MSD), giant number fluctuations, radial
distribution function (RDF), and cluster size distribution. MSD revealed
ballistic behavior at early times and diffusive behavior at later times. The
nature of the transition from ballistic to diffusive regime and the transition
time $\Ttrans$ depend on the region of the $\rm \tau - V$ parameter space. For
MIPS, a smooth transition occurred at the rotational diffusion timescale of
ABPs. In contrast, FIPS and homogeneous phases exhibited a bump after the
transition due to quasi-closed ABP trajectories (Fig.~\ref{fig:Traj}), with
$\Ttrans$ determined by the flow timescale. Flow in FIPS and homogeneous phases control drift velocity and diffusivity, whereas in the MIPS phase,
they are controlled by self-propulsion and rotational diffusion of ABPs. The
exponent $\rm \alpha$ of giant number fluctuation ($\rm {\Delta N_\ell = \rm N_\ell^\alpha}$)
decreased monotonically from $\rm 1$ to $\rm 0.5$ transitioning from MIPS to FIPS to
homogeneous phase. RDF analysis indicated hexagonal closed packing for MIPS,
disordered packing for the homogeneous phase, and an intermediate packing
with both ordered and disordered features for FIPS. Cluster size distribution
showed bimodal behavior for MIPS and FIPS, contrasting with the unimodal
power-law decay of the homogeneous phase.

In this study, hydrodynamic interactions were neglected to
isolate the effect of background flow on MIPS. Previous studies have
demonstrated that hydrodynamic interactions and boundary conditions can induce a
clustered phase for these motile swimmers, a phenomenon also described as
flow-induced phase separation~\cite{thutupalli2018flow}. Crucially, the
aforementioned phase separation relies on hydrodynamic interactions in the
absence of external flow. Eliminating these interactions abolishes phase
separation. Conversely, introducing these hydrodynamic interactions in our
model, depending on whether swimmers are pullers or pushers, alters their
relative motion, impacting both Motility-Induced Phase Separation (MIPS) and
Flow-Induced Phase Separation (FIPS). Specifically, we anticipate a reduced FIPS
and MIPS regime for pullers, and an expanded regime for pushers, compared to our
reported results in Fig.~\ref{fig:comp_phase}. Similar phase separation has
been observed in various systems, including polymers in elongated
flow~\cite{nafar2020flow}, and phase separation of wormlike micelles, a `living'
polymeric system, in flow~\cite{olmsted1999dynamics}. Another interesting
system where we observe phase separation in a fluid medium is photoresponsive
algae under uniform polarized illumination. They swim perpendicular to the
polarization, forming intricate spatial distributions dictated by field
topology~\cite{yang2021controlling}. 

Natural microorganisms reside in fluid environments and often display
non-uniform spatial
distributions~\cite{durham2009disruption,durham2013turbulence}. In active matter
systems, where interactions between constituents play a crucial role, such
non-uniformity can be attributed to MIPS. However, in certain scenarios, these
spatial heterogeneities can originate from external flow conditions
\cite{durham2009disruption,durham2013turbulence}. This study investigates the
combined effect of these mechanisms. To the best of our knowledge, this work
represents the first exploration of the impact of background flow on MIPS. Our
study reveals a novel phase, flow-induced phase separation (FIPS), distinct from
MIPS. The proposed mechanism for FIPS is straightforward and readily adaptable
to experimental settings. We anticipate that this research will stimulate
further investigations into the effects of background flow on MIPS. \\

\section{acknowledgment}
AG is indebted to S Shankar for his tutelage on understanding MIPS. SDP would like to extend her sincere appreciation to S. Ramaswamy for his invaluable discussions.
. SDP and AG would like to thank Kiran Kolluru, K N Banoth and S Mishra for helpful discussions on MIPS. AG acknowledges the funding from SERB-India (grant no. MTR/2022/000232, grant no. CRG/2023/007056-G); DST-India (grant no. DST/NSM/R\&D HPC Applications/2021/05 and grant no. SR/FST/PSI-309 215/2016).

\section{Appendix: $\bar \Lambda$ \& $\uv$} \nonumber

\begin{figure*} 
\includegraphics[width=0.49\linewidth]{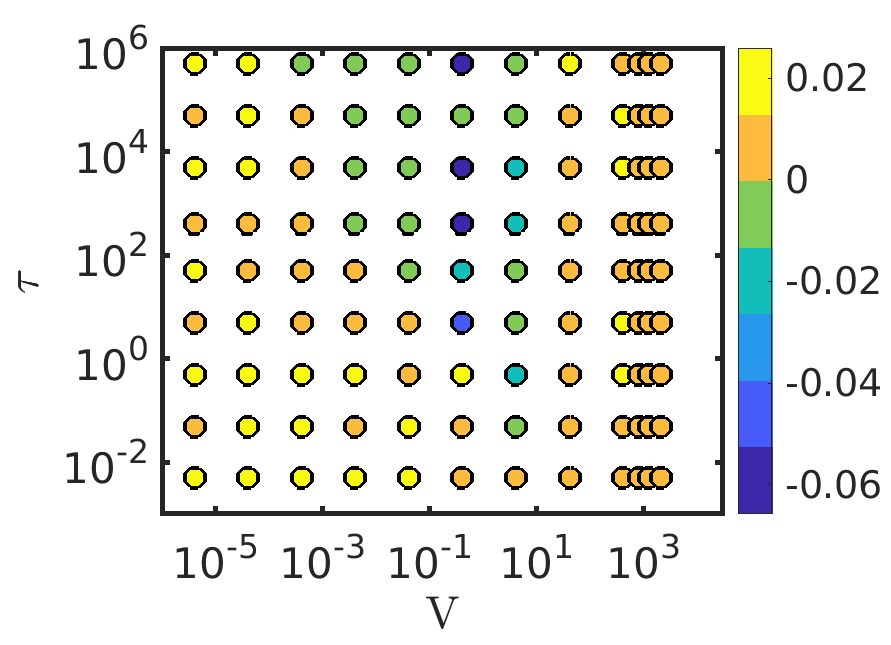} 
\put(-100,145){ { \textcolor{red}{\large \bf (a)} } }
\includegraphics[width=0.49\linewidth]{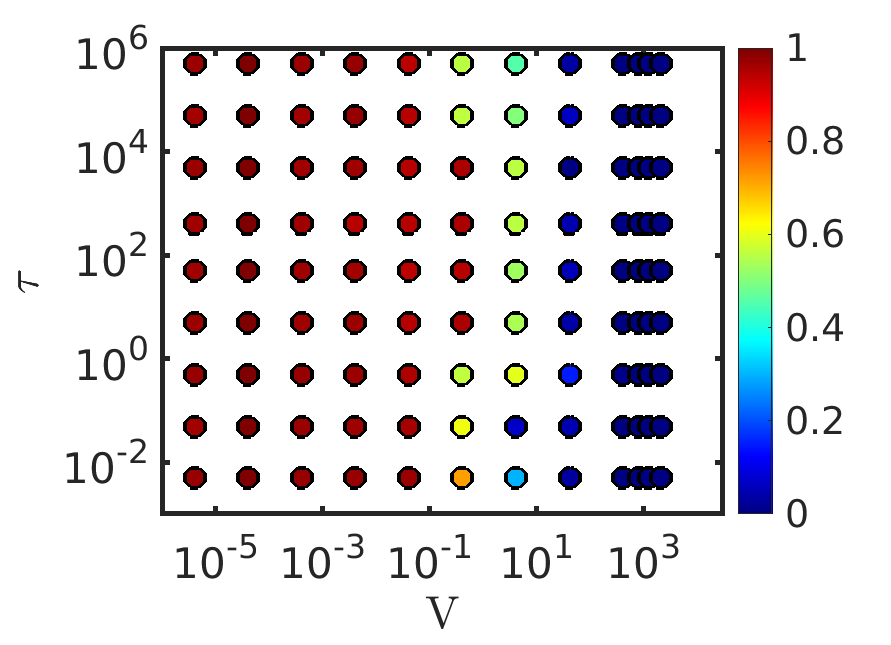} 
\put(-100,145){ { \textcolor{red}{\large \bf (b)} } }
\caption{\label{fig:Lambda}\justifying{(a) $\bar \Lambda$ and (b) $\uv$ values of ABPs in the $\tau$ - $\rm{V}$ plane for $\phi = 0.7$. Colorbars indicate magnitude. The FIPS regions (top middle in the left panel) exhibit negative  $\bar \Lambda$ values, while the MIPS regions (extreme right in the right panel) show $\uv$ values close to zero.}}
\end{figure*}

To distinguish between the three observed phases --- MIPS, homogeneous, and FIPS --- we defined an order parameter, $\OP$, based on two key characteristics. First, in the FIPS phase, active Brownian particles (ABPs) accumulate in strain-dominated regions (regions with large Okubo-Weiss parameter $\Lambda$). This leads to a large negative value for the average Okubo-Weiss parameter,  
$\rm \bar {\Lambda}
\equiv 1/N \sum_{i} \langle \Lambda_i \rangle_t$
 , where $\langle \rangle_t$ denotes the time average in the statistical steady state. In contrast, $\rm \bar {\Lambda}$ is close to zero for both the MIPS and homogeneous phases, as shown in Fig.~\ref{fig:Lambda}(a).

Second, when the background flow velocity, $\rm V$, is much smaller than 1 ($\rm V \ll 1$), the ABP velocities are strongly correlated with the background flow. We quantify this correlation using a normalized velocity correlation, $\uv$, defined as:
\begin{equation}
    \rm \uv \equiv \frac 1 N
\sum_{i} \left \langle \frac{{\bf u}({\bf r}_i,t) \cdot {\bf v}_i} {|{\bf u}({\bf
r}_i,t)| |{\bf v}_i)|} \right \rangle_t. \nonumber
\end{equation} 
We expect $\uv$ to approach 1 when ($\rm V \ll 1$), indicating a strong correlation. As $\rm V$ increases, the correlation weakens, and $\uv$ decreases. Specifically, Fig.~\ref{fig:Lambda}(b) shows that $\uv \approx 1$ for $\rm V \ll 1$, $\uv \approx 0.5$ for $\rm V \sim 1$, and $\uv \approx 0$ for $\rm V \gg 1$.

Hence, by combining the above
information, we can define the following order parameter $\OP$:
\begin{equation}
\OP = \rm \Theta(-\bar
\Lambda) + \Theta(\uv-\epsilon) \nonumber
\end{equation}
where $\rm \epsilon$ is a threshold value that we
have chosen to be $\rm 0.02$ (based on the values from Fig.~\ref{fig:Lambda}(b)) to differentiate the $\uv$ value from MIPS to the rest of the regimes. 

\bibliography{V3MIPS}

\providecommand{\noopsort}[1]{}\providecommand{\singleletter}[1]{#1}%
\begin{thebibliography}{47}%
\makeatletter
\providecommand \@ifxundefined [1]{%
 \@ifx{#1\undefined}
}%
\providecommand \@ifnum [1]{%
 \ifnum #1\expandafter \@firstoftwo
 \else \expandafter \@secondoftwo
 \fi
}%
\providecommand \@ifx [1]{%
 \ifx #1\expandafter \@firstoftwo
 \else \expandafter \@secondoftwo
 \fi
}%
\providecommand \natexlab [1]{#1}%
\providecommand \enquote  [1]{``#1''}%
\providecommand \bibnamefont  [1]{#1}%
\providecommand \bibfnamefont [1]{#1}%
\providecommand \citenamefont [1]{#1}%
\providecommand \href@noop [0]{\@secondoftwo}%
\providecommand \href [0]{\begingroup \@sanitize@url \@href}%
\providecommand \@href[1]{\@@startlink{#1}\@@href}%
\providecommand \@@href[1]{\endgroup#1\@@endlink}%
\providecommand \@sanitize@url [0]{\catcode `\\12\catcode `\$12\catcode
  `\&12\catcode `\#12\catcode `\^12\catcode `\_12\catcode `\%12\relax}%
\providecommand \@@startlink[1]{}%
\providecommand \@@endlink[0]{}%
\providecommand \url  [0]{\begingroup\@sanitize@url \@url }%
\providecommand \@url [1]{\endgroup\@href {#1}{\urlprefix }}%
\providecommand \urlprefix  [0]{URL }%
\providecommand \Eprint [0]{\href }%
\providecommand \doibase [0]{https://doi.org/}%
\providecommand \selectlanguage [0]{\@gobble}%
\providecommand \bibinfo  [0]{\@secondoftwo}%
\providecommand \bibfield  [0]{\@secondoftwo}%
\providecommand \translation [1]{[#1]}%
\providecommand \BibitemOpen [0]{}%
\providecommand \bibitemStop [0]{}%
\providecommand \bibitemNoStop [0]{.\EOS\space}%
\providecommand \EOS [0]{\spacefactor3000\relax}%
\providecommand \BibitemShut  [1]{\csname bibitem#1\endcsname}%
\let\auto@bib@innerbib\@empty
\bibitem [{\citenamefont {Marchetti}\ \emph {et~al.}(2013)\citenamefont
  {Marchetti}, \citenamefont {Joanny}, \citenamefont {Ramaswamy}, \citenamefont
  {Liverpool}, \citenamefont {Prost}, \citenamefont {Rao},\ and\ \citenamefont
  {Simha}}]{marchetti2013hydrodynamics}%
  \BibitemOpen
  \bibfield  {author} {\bibinfo {author} {\bibfnamefont {M.~C.}\ \bibnamefont
  {Marchetti}}, \bibinfo {author} {\bibfnamefont {J.-F.}\ \bibnamefont
  {Joanny}}, \bibinfo {author} {\bibfnamefont {S.}~\bibnamefont {Ramaswamy}},
  \bibinfo {author} {\bibfnamefont {T.~B.}\ \bibnamefont {Liverpool}}, \bibinfo
  {author} {\bibfnamefont {J.}~\bibnamefont {Prost}}, \bibinfo {author}
  {\bibfnamefont {M.}~\bibnamefont {Rao}},\ and\ \bibinfo {author}
  {\bibfnamefont {R.~A.}\ \bibnamefont {Simha}},\ }\bibfield  {title} {\bibinfo
  {title} {Hydrodynamics of soft active matter},\ }\href@noop {} {\bibfield
  {journal} {\bibinfo  {journal} {Reviews of modern physics}\ }\textbf
  {\bibinfo {volume} {85}},\ \bibinfo {pages} {1143} (\bibinfo {year}
  {2013})}\BibitemShut {NoStop}%
\bibitem [{\citenamefont {Bechinger}\ \emph {et~al.}(2016)\citenamefont
  {Bechinger}, \citenamefont {Di~Leonardo}, \citenamefont {L{\"o}wen},
  \citenamefont {Reichhardt}, \citenamefont {Volpe},\ and\ \citenamefont
  {Volpe}}]{bechinger2016active}%
  \BibitemOpen
  \bibfield  {author} {\bibinfo {author} {\bibfnamefont {C.}~\bibnamefont
  {Bechinger}}, \bibinfo {author} {\bibfnamefont {R.}~\bibnamefont
  {Di~Leonardo}}, \bibinfo {author} {\bibfnamefont {H.}~\bibnamefont
  {L{\"o}wen}}, \bibinfo {author} {\bibfnamefont {C.}~\bibnamefont
  {Reichhardt}}, \bibinfo {author} {\bibfnamefont {G.}~\bibnamefont {Volpe}},\
  and\ \bibinfo {author} {\bibfnamefont {G.}~\bibnamefont {Volpe}},\ }\bibfield
   {title} {\bibinfo {title} {Active particles in complex and crowded
  environments},\ }\href@noop {} {\bibfield  {journal} {\bibinfo  {journal}
  {Reviews of Modern Physics}\ }\textbf {\bibinfo {volume} {88}},\ \bibinfo
  {pages} {045006} (\bibinfo {year} {2016})}\BibitemShut {NoStop}%
\bibitem [{\citenamefont {Wu}\ \emph {et~al.}(2009)\citenamefont {Wu},
  \citenamefont {Kaiser}, \citenamefont {Jiang},\ and\ \citenamefont
  {Alber}}]{wu2009periodic}%
  \BibitemOpen
  \bibfield  {author} {\bibinfo {author} {\bibfnamefont {Y.}~\bibnamefont
  {Wu}}, \bibinfo {author} {\bibfnamefont {A.~D.}\ \bibnamefont {Kaiser}},
  \bibinfo {author} {\bibfnamefont {Y.}~\bibnamefont {Jiang}},\ and\ \bibinfo
  {author} {\bibfnamefont {M.~S.}\ \bibnamefont {Alber}},\ }\bibfield  {title}
  {\bibinfo {title} {Periodic reversal of direction allows myxobacteria to
  swarm},\ }\href@noop {} {\bibfield  {journal} {\bibinfo  {journal}
  {Proceedings of the National Academy of Sciences}\ }\textbf {\bibinfo
  {volume} {106}},\ \bibinfo {pages} {1222} (\bibinfo {year}
  {2009})}\BibitemShut {NoStop}%
\bibitem [{\citenamefont {Sokolov}\ \emph {et~al.}(2010)\citenamefont
  {Sokolov}, \citenamefont {Apodaca}, \citenamefont {Grzybowski},\ and\
  \citenamefont {Aranson}}]{sokolov2010swimming}%
  \BibitemOpen
  \bibfield  {author} {\bibinfo {author} {\bibfnamefont {A.}~\bibnamefont
  {Sokolov}}, \bibinfo {author} {\bibfnamefont {M.~M.}\ \bibnamefont
  {Apodaca}}, \bibinfo {author} {\bibfnamefont {B.~A.}\ \bibnamefont
  {Grzybowski}},\ and\ \bibinfo {author} {\bibfnamefont {I.~S.}\ \bibnamefont
  {Aranson}},\ }\bibfield  {title} {\bibinfo {title} {Swimming bacteria power
  microscopic gears},\ }\href@noop {} {\bibfield  {journal} {\bibinfo
  {journal} {Proceedings of the National Academy of Sciences}\ }\textbf
  {\bibinfo {volume} {107}},\ \bibinfo {pages} {969} (\bibinfo {year}
  {2010})}\BibitemShut {NoStop}%
\bibitem [{\citenamefont {Crosato}\ \emph {et~al.}(2018)\citenamefont
  {Crosato}, \citenamefont {Jiang}, \citenamefont {Lecheval}, \citenamefont
  {Lizier}, \citenamefont {Wang}, \citenamefont {Tichit}, \citenamefont
  {Theraulaz},\ and\ \citenamefont {Prokopenko}}]{crosato2018informative}%
  \BibitemOpen
  \bibfield  {author} {\bibinfo {author} {\bibfnamefont {E.}~\bibnamefont
  {Crosato}}, \bibinfo {author} {\bibfnamefont {L.}~\bibnamefont {Jiang}},
  \bibinfo {author} {\bibfnamefont {V.}~\bibnamefont {Lecheval}}, \bibinfo
  {author} {\bibfnamefont {J.~T.}\ \bibnamefont {Lizier}}, \bibinfo {author}
  {\bibfnamefont {X.~R.}\ \bibnamefont {Wang}}, \bibinfo {author}
  {\bibfnamefont {P.}~\bibnamefont {Tichit}}, \bibinfo {author} {\bibfnamefont
  {G.}~\bibnamefont {Theraulaz}},\ and\ \bibinfo {author} {\bibfnamefont
  {M.}~\bibnamefont {Prokopenko}},\ }\bibfield  {title} {\bibinfo {title}
  {Informative and misinformative interactions in a school of fish},\
  }\href@noop {} {\bibfield  {journal} {\bibinfo  {journal} {Swarm
  Intelligence}\ }\textbf {\bibinfo {volume} {12}},\ \bibinfo {pages} {283}
  (\bibinfo {year} {2018})}\BibitemShut {NoStop}%
\bibitem [{\citenamefont {Nagy}\ \emph {et~al.}(2010)\citenamefont {Nagy},
  \citenamefont {{\'A}kos}, \citenamefont {Biro},\ and\ \citenamefont
  {Vicsek}}]{nagy2010hierarchical}%
  \BibitemOpen
  \bibfield  {author} {\bibinfo {author} {\bibfnamefont {M.}~\bibnamefont
  {Nagy}}, \bibinfo {author} {\bibfnamefont {Z.}~\bibnamefont {{\'A}kos}},
  \bibinfo {author} {\bibfnamefont {D.}~\bibnamefont {Biro}},\ and\ \bibinfo
  {author} {\bibfnamefont {T.}~\bibnamefont {Vicsek}},\ }\bibfield  {title}
  {\bibinfo {title} {Hierarchical group dynamics in pigeon flocks},\
  }\href@noop {} {\bibfield  {journal} {\bibinfo  {journal} {Nature}\ }\textbf
  {\bibinfo {volume} {464}},\ \bibinfo {pages} {890} (\bibinfo {year}
  {2010})}\BibitemShut {NoStop}%
\bibitem [{\citenamefont {Mora}\ \emph {et~al.}(2016)\citenamefont {Mora},
  \citenamefont {Walczak}, \citenamefont {Del~Castello}, \citenamefont
  {Ginelli}, \citenamefont {Melillo}, \citenamefont {Parisi}, \citenamefont
  {Viale}, \citenamefont {Cavagna},\ and\ \citenamefont
  {Giardina}}]{mora2016local}%
  \BibitemOpen
  \bibfield  {author} {\bibinfo {author} {\bibfnamefont {T.}~\bibnamefont
  {Mora}}, \bibinfo {author} {\bibfnamefont {A.~M.}\ \bibnamefont {Walczak}},
  \bibinfo {author} {\bibfnamefont {L.}~\bibnamefont {Del~Castello}}, \bibinfo
  {author} {\bibfnamefont {F.}~\bibnamefont {Ginelli}}, \bibinfo {author}
  {\bibfnamefont {S.}~\bibnamefont {Melillo}}, \bibinfo {author} {\bibfnamefont
  {L.}~\bibnamefont {Parisi}}, \bibinfo {author} {\bibfnamefont
  {M.}~\bibnamefont {Viale}}, \bibinfo {author} {\bibfnamefont
  {A.}~\bibnamefont {Cavagna}},\ and\ \bibinfo {author} {\bibfnamefont
  {I.}~\bibnamefont {Giardina}},\ }\bibfield  {title} {\bibinfo {title} {Local
  equilibrium in bird flocks},\ }\href@noop {} {\bibfield  {journal} {\bibinfo
  {journal} {Nature physics}\ }\textbf {\bibinfo {volume} {12}},\ \bibinfo
  {pages} {1153} (\bibinfo {year} {2016})}\BibitemShut {NoStop}%
\bibitem [{\citenamefont {Hueschen}\ \emph {et~al.}(2023)\citenamefont
  {Hueschen}, \citenamefont {Dunn},\ and\ \citenamefont
  {Phillips}}]{hueschen2023wildebeest}%
  \BibitemOpen
  \bibfield  {author} {\bibinfo {author} {\bibfnamefont {C.~L.}\ \bibnamefont
  {Hueschen}}, \bibinfo {author} {\bibfnamefont {A.~R.}\ \bibnamefont {Dunn}},\
  and\ \bibinfo {author} {\bibfnamefont {R.}~\bibnamefont {Phillips}},\
  }\bibfield  {title} {\bibinfo {title} {Wildebeest herds on rolling hills:
  Flocking on arbitrary curved surfaces},\ }\href@noop {} {\bibfield  {journal}
  {\bibinfo  {journal} {Physical Review E}\ }\textbf {\bibinfo {volume}
  {108}},\ \bibinfo {pages} {024610} (\bibinfo {year} {2023})}\BibitemShut
  {NoStop}%
\bibitem [{\citenamefont {Garcimart{\'\i}n}\ \emph {et~al.}(2015)\citenamefont
  {Garcimart{\'\i}n}, \citenamefont {Pastor}, \citenamefont {Ferrer},
  \citenamefont {Ramos}, \citenamefont {Mart{\'\i}n-G{\'o}mez},\ and\
  \citenamefont {Zuriguel}}]{garcimartin2015flow}%
  \BibitemOpen
  \bibfield  {author} {\bibinfo {author} {\bibfnamefont {A.}~\bibnamefont
  {Garcimart{\'\i}n}}, \bibinfo {author} {\bibfnamefont {J.}~\bibnamefont
  {Pastor}}, \bibinfo {author} {\bibfnamefont {L.}~\bibnamefont {Ferrer}},
  \bibinfo {author} {\bibfnamefont {J.}~\bibnamefont {Ramos}}, \bibinfo
  {author} {\bibfnamefont {C.}~\bibnamefont {Mart{\'\i}n-G{\'o}mez}},\ and\
  \bibinfo {author} {\bibfnamefont {I.}~\bibnamefont {Zuriguel}},\ }\bibfield
  {title} {\bibinfo {title} {Flow and clogging of a sheep herd passing through
  a bottleneck},\ }\href@noop {} {\bibfield  {journal} {\bibinfo  {journal}
  {Physical Review E}\ }\textbf {\bibinfo {volume} {91}},\ \bibinfo {pages}
  {022808} (\bibinfo {year} {2015})}\BibitemShut {NoStop}%
\bibitem [{\citenamefont {Bellomo}\ \emph {et~al.}(2016)\citenamefont
  {Bellomo}, \citenamefont {Clarke}, \citenamefont {Gibelli}, \citenamefont
  {Townsend},\ and\ \citenamefont {Vreugdenhil}}]{bellomo2016human}%
  \BibitemOpen
  \bibfield  {author} {\bibinfo {author} {\bibfnamefont {N.}~\bibnamefont
  {Bellomo}}, \bibinfo {author} {\bibfnamefont {D.}~\bibnamefont {Clarke}},
  \bibinfo {author} {\bibfnamefont {L.}~\bibnamefont {Gibelli}}, \bibinfo
  {author} {\bibfnamefont {P.}~\bibnamefont {Townsend}},\ and\ \bibinfo
  {author} {\bibfnamefont {B.}~\bibnamefont {Vreugdenhil}},\ }\bibfield
  {title} {\bibinfo {title} {Human behaviours in evacuation crowd dynamics:
  From modelling to “big data” toward crisis management},\ }\href@noop {}
  {\bibfield  {journal} {\bibinfo  {journal} {Physics of life reviews}\
  }\textbf {\bibinfo {volume} {18}},\ \bibinfo {pages} {1} (\bibinfo {year}
  {2016})}\BibitemShut {NoStop}%
\bibitem [{\citenamefont {Jiang}\ \emph {et~al.}(2010)\citenamefont {Jiang},
  \citenamefont {Yoshinaga},\ and\ \citenamefont {Sano}}]{jiang2010active}%
  \BibitemOpen
  \bibfield  {author} {\bibinfo {author} {\bibfnamefont {H.-R.}\ \bibnamefont
  {Jiang}}, \bibinfo {author} {\bibfnamefont {N.}~\bibnamefont {Yoshinaga}},\
  and\ \bibinfo {author} {\bibfnamefont {M.}~\bibnamefont {Sano}},\ }\bibfield
  {title} {\bibinfo {title} {Active motion of a janus particle by
  self-thermophoresis in a defocused laser beam},\ }\href@noop {} {\bibfield
  {journal} {\bibinfo  {journal} {Physical review letters}\ }\textbf {\bibinfo
  {volume} {105}},\ \bibinfo {pages} {268302} (\bibinfo {year}
  {2010})}\BibitemShut {NoStop}%
\bibitem [{\citenamefont {Narayan}\ \emph {et~al.}(2007)\citenamefont
  {Narayan}, \citenamefont {Ramaswamy},\ and\ \citenamefont
  {Menon}}]{Narayan2007}%
  \BibitemOpen
  \bibfield  {author} {\bibinfo {author} {\bibfnamefont {V.}~\bibnamefont
  {Narayan}}, \bibinfo {author} {\bibfnamefont {S.}~\bibnamefont {Ramaswamy}},\
  and\ \bibinfo {author} {\bibfnamefont {N.}~\bibnamefont {Menon}},\ }\bibfield
   {title} {\bibinfo {title} {{Long-lived giant number fluctuations in a
  swarming granular nematic}},\ }\href@noop {} {\bibfield  {journal} {\bibinfo
  {journal} {Science}\ }\textbf {\bibinfo {volume} {317}},\ \bibinfo {pages}
  {105} (\bibinfo {year} {2007})}\BibitemShut {NoStop}%
\bibitem [{\citenamefont {Dey}\ \emph {et~al.}(2022)\citenamefont {Dey},
  \citenamefont {Buness}, \citenamefont {Hokmabad}, \citenamefont {Jin},\ and\
  \citenamefont {Maass}}]{dey2022oscillatory}%
  \BibitemOpen
  \bibfield  {author} {\bibinfo {author} {\bibfnamefont {R.}~\bibnamefont
  {Dey}}, \bibinfo {author} {\bibfnamefont {C.~M.}\ \bibnamefont {Buness}},
  \bibinfo {author} {\bibfnamefont {B.~V.}\ \bibnamefont {Hokmabad}}, \bibinfo
  {author} {\bibfnamefont {C.}~\bibnamefont {Jin}},\ and\ \bibinfo {author}
  {\bibfnamefont {C.~C.}\ \bibnamefont {Maass}},\ }\bibfield  {title} {\bibinfo
  {title} {Oscillatory rheotaxis of artificial swimmers in microchannels},\
  }\href@noop {} {\bibfield  {journal} {\bibinfo  {journal} {Nature
  communications}\ }\textbf {\bibinfo {volume} {13}},\ \bibinfo {pages} {2952}
  (\bibinfo {year} {2022})}\BibitemShut {NoStop}%
\bibitem [{\citenamefont {O’Keeffe}\ \emph {et~al.}(2017)\citenamefont
  {O’Keeffe}, \citenamefont {Hong},\ and\ \citenamefont
  {Strogatz}}]{o2017oscillators}%
  \BibitemOpen
  \bibfield  {author} {\bibinfo {author} {\bibfnamefont {K.~P.}\ \bibnamefont
  {O’Keeffe}}, \bibinfo {author} {\bibfnamefont {H.}~\bibnamefont {Hong}},\
  and\ \bibinfo {author} {\bibfnamefont {S.~H.}\ \bibnamefont {Strogatz}},\
  }\bibfield  {title} {\bibinfo {title} {Oscillators that sync and swarm},\
  }\href@noop {} {\bibfield  {journal} {\bibinfo  {journal} {Nature
  communications}\ }\textbf {\bibinfo {volume} {8}},\ \bibinfo {pages} {1504}
  (\bibinfo {year} {2017})}\BibitemShut {NoStop}%
\bibitem [{\citenamefont {Paramanick}\ \emph {et~al.}(2024)\citenamefont
  {Paramanick}, \citenamefont {Pal}, \citenamefont {Soni},\ and\ \citenamefont
  {Kumar}}]{paramanick2024programming}%
  \BibitemOpen
  \bibfield  {author} {\bibinfo {author} {\bibfnamefont {S.}~\bibnamefont
  {Paramanick}}, \bibinfo {author} {\bibfnamefont {A.}~\bibnamefont {Pal}},
  \bibinfo {author} {\bibfnamefont {H.}~\bibnamefont {Soni}},\ and\ \bibinfo
  {author} {\bibfnamefont {N.}~\bibnamefont {Kumar}},\ }\bibfield  {title}
  {\bibinfo {title} {Programming tunable active dynamics in a self-propelled
  robot},\ }\href@noop {} {\bibfield  {journal} {\bibinfo  {journal} {The
  European Physical Journal E}\ }\textbf {\bibinfo {volume} {47}},\ \bibinfo
  {pages} {34} (\bibinfo {year} {2024})}\BibitemShut {NoStop}%
\bibitem [{\citenamefont {Redner}\ \emph {et~al.}(2013)\citenamefont {Redner},
  \citenamefont {Hagan},\ and\ \citenamefont {Baskaran}}]{redner2013structure}%
  \BibitemOpen
  \bibfield  {author} {\bibinfo {author} {\bibfnamefont {G.~S.}\ \bibnamefont
  {Redner}}, \bibinfo {author} {\bibfnamefont {M.~F.}\ \bibnamefont {Hagan}},\
  and\ \bibinfo {author} {\bibfnamefont {A.}~\bibnamefont {Baskaran}},\
  }\bibfield  {title} {\bibinfo {title} {Structure and dynamics of a
  phase-separating active colloidal fluid},\ }\href
  {https://doi.org/https://doi.org/10.1103/PhysRevLett.110.055701} {\bibfield
  {journal} {\bibinfo  {journal} {Physical Review Letters}\ }\textbf {\bibinfo
  {volume} {110}},\ \bibinfo {pages} {055701} (\bibinfo {year}
  {2013})}\BibitemShut {NoStop}%
\bibitem [{\citenamefont {Fily}\ and\ \citenamefont
  {Marchetti}(2012)}]{Fily2012}%
  \BibitemOpen
  \bibfield  {author} {\bibinfo {author} {\bibfnamefont {Y.}~\bibnamefont
  {Fily}}\ and\ \bibinfo {author} {\bibfnamefont {M.~C.}\ \bibnamefont
  {Marchetti}},\ }\bibfield  {title} {\bibinfo {title} {{Athermal phase
  separation of self-propelled particles with no alignment}},\ }\href
  {https://doi.org/10.1103/PhysRevLett.108.235702} {\bibfield  {journal}
  {\bibinfo  {journal} {Physical Review Letters}\ }\textbf {\bibinfo {volume}
  {108}},\ \bibinfo {pages} {1} (\bibinfo {year} {2012})}\BibitemShut {NoStop}%
\bibitem [{\citenamefont {Vicsek}\ \emph {et~al.}(1995)\citenamefont {Vicsek},
  \citenamefont {Czir{\'o}k}, \citenamefont {Ben-Jacob}, \citenamefont
  {Cohen},\ and\ \citenamefont {Shochet}}]{vicsek1995novel}%
  \BibitemOpen
  \bibfield  {author} {\bibinfo {author} {\bibfnamefont {T.}~\bibnamefont
  {Vicsek}}, \bibinfo {author} {\bibfnamefont {A.}~\bibnamefont {Czir{\'o}k}},
  \bibinfo {author} {\bibfnamefont {E.}~\bibnamefont {Ben-Jacob}}, \bibinfo
  {author} {\bibfnamefont {I.}~\bibnamefont {Cohen}},\ and\ \bibinfo {author}
  {\bibfnamefont {O.}~\bibnamefont {Shochet}},\ }\bibfield  {title} {\bibinfo
  {title} {Novel type of phase transition in a system of self-driven
  particles},\ }\href@noop {} {\bibfield  {journal} {\bibinfo  {journal}
  {Physical review letters}\ }\textbf {\bibinfo {volume} {75}},\ \bibinfo
  {pages} {1226} (\bibinfo {year} {1995})}\BibitemShut {NoStop}%
\bibitem [{\citenamefont {Martin}\ and\ \citenamefont
  {de~Pirey}(2021)}]{martin2021aoup}%
  \BibitemOpen
  \bibfield  {author} {\bibinfo {author} {\bibfnamefont {D.}~\bibnamefont
  {Martin}}\ and\ \bibinfo {author} {\bibfnamefont {T.~A.}\ \bibnamefont
  {de~Pirey}},\ }\bibfield  {title} {\bibinfo {title} {Aoup in the presence of
  brownian noise: a perturbative approach},\ }\href@noop {} {\bibfield
  {journal} {\bibinfo  {journal} {Journal of Statistical Mechanics: Theory and
  Experiment}\ }\textbf {\bibinfo {volume} {2021}},\ \bibinfo {pages} {043205}
  (\bibinfo {year} {2021})}\BibitemShut {NoStop}%
\bibitem [{\citenamefont {Sanoria}\ \emph {et~al.}(2021)\citenamefont
  {Sanoria}, \citenamefont {Chelakkot},\ and\ \citenamefont
  {Nandi}}]{sanoria2021influence}%
  \BibitemOpen
  \bibfield  {author} {\bibinfo {author} {\bibfnamefont {M.}~\bibnamefont
  {Sanoria}}, \bibinfo {author} {\bibfnamefont {R.}~\bibnamefont {Chelakkot}},\
  and\ \bibinfo {author} {\bibfnamefont {A.}~\bibnamefont {Nandi}},\ }\bibfield
   {title} {\bibinfo {title} {Influence of interaction softness on phase
  separation of active particles},\ }\href@noop {} {\bibfield  {journal}
  {\bibinfo  {journal} {Physical Review E}\ }\textbf {\bibinfo {volume}
  {103}},\ \bibinfo {pages} {052605} (\bibinfo {year} {2021})}\BibitemShut
  {NoStop}%
\bibitem [{\citenamefont {Bertrand}\ \emph {et~al.}(2018)\citenamefont
  {Bertrand}, \citenamefont {Zhao}, \citenamefont {B{\'e}nichou}, \citenamefont
  {Tailleur},\ and\ \citenamefont {Voituriez}}]{bertrand2018optimized}%
  \BibitemOpen
  \bibfield  {author} {\bibinfo {author} {\bibfnamefont {T.}~\bibnamefont
  {Bertrand}}, \bibinfo {author} {\bibfnamefont {Y.}~\bibnamefont {Zhao}},
  \bibinfo {author} {\bibfnamefont {O.}~\bibnamefont {B{\'e}nichou}}, \bibinfo
  {author} {\bibfnamefont {J.}~\bibnamefont {Tailleur}},\ and\ \bibinfo
  {author} {\bibfnamefont {R.}~\bibnamefont {Voituriez}},\ }\bibfield  {title}
  {\bibinfo {title} {Optimized diffusion of run-and-tumble particles in crowded
  environments},\ }\href@noop {} {\bibfield  {journal} {\bibinfo  {journal}
  {Physical Review Letters}\ }\textbf {\bibinfo {volume} {120}},\ \bibinfo
  {pages} {198103} (\bibinfo {year} {2018})}\BibitemShut {NoStop}%
\bibitem [{\citenamefont {Santra}\ \emph {et~al.}(2020)\citenamefont {Santra},
  \citenamefont {Basu},\ and\ \citenamefont {Sabhapandit}}]{santra2020run}%
  \BibitemOpen
  \bibfield  {author} {\bibinfo {author} {\bibfnamefont {I.}~\bibnamefont
  {Santra}}, \bibinfo {author} {\bibfnamefont {U.}~\bibnamefont {Basu}},\ and\
  \bibinfo {author} {\bibfnamefont {S.}~\bibnamefont {Sabhapandit}},\
  }\bibfield  {title} {\bibinfo {title} {Run-and-tumble particles in two
  dimensions: Marginal position distributions},\ }\href@noop {} {\bibfield
  {journal} {\bibinfo  {journal} {Physical Review E}\ }\textbf {\bibinfo
  {volume} {101}},\ \bibinfo {pages} {062120} (\bibinfo {year}
  {2020})}\BibitemShut {NoStop}%
\bibitem [{\citenamefont {Toner}\ \emph {et~al.}(2005)\citenamefont {Toner},
  \citenamefont {Tu},\ and\ \citenamefont
  {Ramaswamy}}]{toner2005hydrodynamics}%
  \BibitemOpen
  \bibfield  {author} {\bibinfo {author} {\bibfnamefont {J.}~\bibnamefont
  {Toner}}, \bibinfo {author} {\bibfnamefont {Y.}~\bibnamefont {Tu}},\ and\
  \bibinfo {author} {\bibfnamefont {S.}~\bibnamefont {Ramaswamy}},\ }\bibfield
  {title} {\bibinfo {title} {Hydrodynamics and phases of flocks},\ }\href@noop
  {} {\bibfield  {journal} {\bibinfo  {journal} {Annals of Physics}\ }\textbf
  {\bibinfo {volume} {318}},\ \bibinfo {pages} {170} (\bibinfo {year}
  {2005})}\BibitemShut {NoStop}%
\bibitem [{\citenamefont {Sanoria}\ \emph {et~al.}(2022)\citenamefont
  {Sanoria}, \citenamefont {Chelakkot},\ and\ \citenamefont
  {Nandi}}]{sanoria2022percolation}%
  \BibitemOpen
  \bibfield  {author} {\bibinfo {author} {\bibfnamefont {M.}~\bibnamefont
  {Sanoria}}, \bibinfo {author} {\bibfnamefont {R.}~\bibnamefont {Chelakkot}},\
  and\ \bibinfo {author} {\bibfnamefont {A.}~\bibnamefont {Nandi}},\ }\bibfield
   {title} {\bibinfo {title} {Percolation transition in phase-separating active
  fluid},\ }\href@noop {} {\bibfield  {journal} {\bibinfo  {journal} {Physical
  Review E}\ }\textbf {\bibinfo {volume} {106}},\ \bibinfo {pages} {034605}
  (\bibinfo {year} {2022})}\BibitemShut {NoStop}%
\bibitem [{\citenamefont {Dolai}\ \emph {et~al.}(2018)\citenamefont {Dolai},
  \citenamefont {Simha},\ and\ \citenamefont {Mishra}}]{dolai2018phase}%
  \BibitemOpen
  \bibfield  {author} {\bibinfo {author} {\bibfnamefont {P.}~\bibnamefont
  {Dolai}}, \bibinfo {author} {\bibfnamefont {A.}~\bibnamefont {Simha}},\ and\
  \bibinfo {author} {\bibfnamefont {S.}~\bibnamefont {Mishra}},\ }\bibfield
  {title} {\bibinfo {title} {Phase separation in binary mixtures of active and
  passive particles},\ }\href@noop {} {\bibfield  {journal} {\bibinfo
  {journal} {Soft Matter}\ }\textbf {\bibinfo {volume} {14}},\ \bibinfo {pages}
  {6137} (\bibinfo {year} {2018})}\BibitemShut {NoStop}%
\bibitem [{\citenamefont {Gutierrez-Castillo}\ and\ \citenamefont
  {Thomases}(2019)}]{gutierrez2019proper}%
  \BibitemOpen
  \bibfield  {author} {\bibinfo {author} {\bibfnamefont {P.}~\bibnamefont
  {Gutierrez-Castillo}}\ and\ \bibinfo {author} {\bibfnamefont
  {B.}~\bibnamefont {Thomases}},\ }\bibfield  {title} {\bibinfo {title} {Proper
  orthogonal decomposition (pod) of the flow dynamics for a viscoelastic fluid
  in a four-roll mill geometry at the stokes limit},\ }\href@noop {} {\bibfield
   {journal} {\bibinfo  {journal} {Journal of Non-Newtonian Fluid Mechanics}\
  }\textbf {\bibinfo {volume} {264}},\ \bibinfo {pages} {48} (\bibinfo {year}
  {2019})}\BibitemShut {NoStop}%
\bibitem [{\citenamefont {Lee}\ \emph {et~al.}(2007)\citenamefont {Lee},
  \citenamefont {Dylla-Spears}, \citenamefont {Teclemariam},\ and\
  \citenamefont {Muller}}]{lee2007microfluidic}%
  \BibitemOpen
  \bibfield  {author} {\bibinfo {author} {\bibfnamefont {J.~S.}\ \bibnamefont
  {Lee}}, \bibinfo {author} {\bibfnamefont {R.}~\bibnamefont {Dylla-Spears}},
  \bibinfo {author} {\bibfnamefont {N.~P.}\ \bibnamefont {Teclemariam}},\ and\
  \bibinfo {author} {\bibfnamefont {S.~J.}\ \bibnamefont {Muller}},\ }\bibfield
   {title} {\bibinfo {title} {Microfluidic four-roll mill for all flow types},\
  }\href@noop {} {\bibfield  {journal} {\bibinfo  {journal} {Applied physics
  letters}\ }\textbf {\bibinfo {volume} {90}},\ \bibinfo {pages} {074103}
  (\bibinfo {year} {2007})}\BibitemShut {NoStop}%
\bibitem [{\citenamefont {Solomon}\ and\ \citenamefont
  {Mezi{\'c}}(2003)}]{solomon2003uniform}%
  \BibitemOpen
  \bibfield  {author} {\bibinfo {author} {\bibfnamefont {T.}~\bibnamefont
  {Solomon}}\ and\ \bibinfo {author} {\bibfnamefont {I.}~\bibnamefont
  {Mezi{\'c}}},\ }\bibfield  {title} {\bibinfo {title} {Uniform resonant
  chaotic mixing in fluid flows},\ }\href@noop {} {\bibfield  {journal}
  {\bibinfo  {journal} {Nature}\ }\textbf {\bibinfo {volume} {425}},\ \bibinfo
  {pages} {376} (\bibinfo {year} {2003})}\BibitemShut {NoStop}%
\bibitem [{\citenamefont {Solomon}\ and\ \citenamefont
  {Gollub}(1988)}]{solomon1988chaotic}%
  \BibitemOpen
  \bibfield  {author} {\bibinfo {author} {\bibfnamefont {T.}~\bibnamefont
  {Solomon}}\ and\ \bibinfo {author} {\bibfnamefont {J.~P.}\ \bibnamefont
  {Gollub}},\ }\bibfield  {title} {\bibinfo {title} {Chaotic particle transport
  in time-dependent rayleigh-b{\'e}nard convection},\ }\href@noop {} {\bibfield
   {journal} {\bibinfo  {journal} {Physical Review A}\ }\textbf {\bibinfo
  {volume} {38}},\ \bibinfo {pages} {6280} (\bibinfo {year}
  {1988})}\BibitemShut {NoStop}%
\bibitem [{\citenamefont {Pandit}\ \emph {et~al.}(2017)\citenamefont {Pandit},
  \citenamefont {Banerjee}, \citenamefont {Bhatnagar}, \citenamefont {Brachet},
  \citenamefont {Gupta}, \citenamefont {Mitra}, \citenamefont {Pal},
  \citenamefont {Perlekar}, \citenamefont {Ray}, \citenamefont {Shukla} \emph
  {et~al.}}]{pandit2017overview}%
  \BibitemOpen
  \bibfield  {author} {\bibinfo {author} {\bibfnamefont {R.}~\bibnamefont
  {Pandit}}, \bibinfo {author} {\bibfnamefont {D.}~\bibnamefont {Banerjee}},
  \bibinfo {author} {\bibfnamefont {A.}~\bibnamefont {Bhatnagar}}, \bibinfo
  {author} {\bibfnamefont {M.}~\bibnamefont {Brachet}}, \bibinfo {author}
  {\bibfnamefont {A.}~\bibnamefont {Gupta}}, \bibinfo {author} {\bibfnamefont
  {D.}~\bibnamefont {Mitra}}, \bibinfo {author} {\bibfnamefont
  {N.}~\bibnamefont {Pal}}, \bibinfo {author} {\bibfnamefont {P.}~\bibnamefont
  {Perlekar}}, \bibinfo {author} {\bibfnamefont {S.~S.}\ \bibnamefont {Ray}},
  \bibinfo {author} {\bibfnamefont {V.}~\bibnamefont {Shukla}}, \emph
  {et~al.},\ }\bibfield  {title} {\bibinfo {title} {An overview of the
  statistical properties of two-dimensional turbulence in fluids with
  particles, conducting fluids, fluids with polymer additives, binary-fluid
  mixtures, and superfluids},\ }\href@noop {} {\bibfield  {journal} {\bibinfo
  {journal} {Physics of fluids}\ }\textbf {\bibinfo {volume} {29}},\ \bibinfo
  {pages} {111112} (\bibinfo {year} {2017})}\BibitemShut {NoStop}%
\bibitem [{\citenamefont {Gupta}\ \emph {et~al.}(2015)\citenamefont {Gupta},
  \citenamefont {Perlekar},\ and\ \citenamefont {Pandit}}]{gupta2015two}%
  \BibitemOpen
  \bibfield  {author} {\bibinfo {author} {\bibfnamefont {A.}~\bibnamefont
  {Gupta}}, \bibinfo {author} {\bibfnamefont {P.}~\bibnamefont {Perlekar}},\
  and\ \bibinfo {author} {\bibfnamefont {R.}~\bibnamefont {Pandit}},\
  }\bibfield  {title} {\bibinfo {title} {Two-dimensional homogeneous isotropic
  fluid turbulence with polymer additives},\ }\href@noop {} {\bibfield
  {journal} {\bibinfo  {journal} {Physical Review E}\ }\textbf {\bibinfo
  {volume} {91}},\ \bibinfo {pages} {033013} (\bibinfo {year}
  {2015})}\BibitemShut {NoStop}%
\bibitem [{\citenamefont {Gupta}\ \emph {et~al.}(2014)\citenamefont {Gupta},
  \citenamefont {Vincenzi},\ and\ \citenamefont
  {Pandit}}]{gupta2014elliptical}%
  \BibitemOpen
  \bibfield  {author} {\bibinfo {author} {\bibfnamefont {A.}~\bibnamefont
  {Gupta}}, \bibinfo {author} {\bibfnamefont {D.}~\bibnamefont {Vincenzi}},\
  and\ \bibinfo {author} {\bibfnamefont {R.}~\bibnamefont {Pandit}},\
  }\bibfield  {title} {\bibinfo {title} {Elliptical tracers in two-dimensional,
  homogeneous, isotropic fluid turbulence: The statistics of alignment,
  rotation, and nematic order},\ }\href@noop {} {\bibfield  {journal} {\bibinfo
   {journal} {Physical Review E}\ }\textbf {\bibinfo {volume} {89}},\ \bibinfo
  {pages} {021001} (\bibinfo {year} {2014})}\BibitemShut {NoStop}%
\bibitem [{\citenamefont {Torney}\ and\ \citenamefont
  {Neufeld}(2007)}]{torney2007transport}%
  \BibitemOpen
  \bibfield  {author} {\bibinfo {author} {\bibfnamefont {C.}~\bibnamefont
  {Torney}}\ and\ \bibinfo {author} {\bibfnamefont {Z.}~\bibnamefont
  {Neufeld}},\ }\bibfield  {title} {\bibinfo {title} {Transport and aggregation
  of self-propelled particles in fluid flows},\ }\href@noop {} {\bibfield
  {journal} {\bibinfo  {journal} {Physical review letters}\ }\textbf {\bibinfo
  {volume} {99}},\ \bibinfo {pages} {078101} (\bibinfo {year}
  {2007})}\BibitemShut {NoStop}%
\bibitem [{\citenamefont {Malakar}\ \emph {et~al.}(2020)\citenamefont
  {Malakar}, \citenamefont {Das}, \citenamefont {Kundu}, \citenamefont
  {Kumar},\ and\ \citenamefont {Dhar}}]{malakar2020steady}%
  \BibitemOpen
  \bibfield  {author} {\bibinfo {author} {\bibfnamefont {K.}~\bibnamefont
  {Malakar}}, \bibinfo {author} {\bibfnamefont {A.}~\bibnamefont {Das}},
  \bibinfo {author} {\bibfnamefont {A.}~\bibnamefont {Kundu}}, \bibinfo
  {author} {\bibfnamefont {K.~V.}\ \bibnamefont {Kumar}},\ and\ \bibinfo
  {author} {\bibfnamefont {A.}~\bibnamefont {Dhar}},\ }\bibfield  {title}
  {\bibinfo {title} {Steady state of an active brownian particle in a
  two-dimensional harmonic trap},\ }\href@noop {} {\bibfield  {journal}
  {\bibinfo  {journal} {Physical Review E}\ }\textbf {\bibinfo {volume}
  {101}},\ \bibinfo {pages} {022610} (\bibinfo {year} {2020})}\BibitemShut
  {NoStop}%
\bibitem [{\citenamefont {Janzen}\ and\ \citenamefont
  {Janssen}(2022)}]{janzen2022aging}%
  \BibitemOpen
  \bibfield  {author} {\bibinfo {author} {\bibfnamefont {G.}~\bibnamefont
  {Janzen}}\ and\ \bibinfo {author} {\bibfnamefont {L.~M.}\ \bibnamefont
  {Janssen}},\ }\bibfield  {title} {\bibinfo {title} {Aging in thermal active
  glasses},\ }\href@noop {} {\bibfield  {journal} {\bibinfo  {journal}
  {Physical Review Research}\ }\textbf {\bibinfo {volume} {4}},\ \bibinfo
  {pages} {L012038} (\bibinfo {year} {2022})}\BibitemShut {NoStop}%
\bibitem [{\citenamefont {Perlekar}\ \emph {et~al.}(2011)\citenamefont
  {Perlekar}, \citenamefont {Ray}, \citenamefont {Mitra},\ and\ \citenamefont
  {Pandit}}]{perlekar2011persistence}%
  \BibitemOpen
  \bibfield  {author} {\bibinfo {author} {\bibfnamefont {P.}~\bibnamefont
  {Perlekar}}, \bibinfo {author} {\bibfnamefont {S.~S.}\ \bibnamefont {Ray}},
  \bibinfo {author} {\bibfnamefont {D.}~\bibnamefont {Mitra}},\ and\ \bibinfo
  {author} {\bibfnamefont {R.}~\bibnamefont {Pandit}},\ }\bibfield  {title}
  {\bibinfo {title} {Persistence problem in two-dimensional fluid turbulence},\
  }\href@noop {} {\bibfield  {journal} {\bibinfo  {journal} {Physical review
  letters}\ }\textbf {\bibinfo {volume} {106}},\ \bibinfo {pages} {054501}
  (\bibinfo {year} {2011})}\BibitemShut {NoStop}%
\bibitem [{\citenamefont {Caprini}\ \emph {et~al.}(2020)\citenamefont
  {Caprini}, \citenamefont {Cecconi}, \citenamefont {Puglisi},\ and\
  \citenamefont {Sarracino}}]{caprini2020diffusion}%
  \BibitemOpen
  \bibfield  {author} {\bibinfo {author} {\bibfnamefont {L.}~\bibnamefont
  {Caprini}}, \bibinfo {author} {\bibfnamefont {F.}~\bibnamefont {Cecconi}},
  \bibinfo {author} {\bibfnamefont {A.}~\bibnamefont {Puglisi}},\ and\ \bibinfo
  {author} {\bibfnamefont {A.}~\bibnamefont {Sarracino}},\ }\bibfield  {title}
  {\bibinfo {title} {Diffusion properties of self-propelled particles in
  cellular flows},\ }\href@noop {} {\bibfield  {journal} {\bibinfo  {journal}
  {Soft Matter}\ }\textbf {\bibinfo {volume} {16}},\ \bibinfo {pages} {5431}
  (\bibinfo {year} {2020})}\BibitemShut {NoStop}%
\bibitem [{\citenamefont {Henkes}\ \emph {et~al.}(2011)\citenamefont {Henkes},
  \citenamefont {Fily},\ and\ \citenamefont {Marchetti}}]{henkes2011active}%
  \BibitemOpen
  \bibfield  {author} {\bibinfo {author} {\bibfnamefont {S.}~\bibnamefont
  {Henkes}}, \bibinfo {author} {\bibfnamefont {Y.}~\bibnamefont {Fily}},\ and\
  \bibinfo {author} {\bibfnamefont {M.~C.}\ \bibnamefont {Marchetti}},\
  }\bibfield  {title} {\bibinfo {title} {Active jamming: Self-propelled soft
  particles at high density},\ }\href@noop {} {\bibfield  {journal} {\bibinfo
  {journal} {Physical Review E}\ }\textbf {\bibinfo {volume} {84}},\ \bibinfo
  {pages} {040301} (\bibinfo {year} {2011})}\BibitemShut {NoStop}%
\bibitem [{\citenamefont {Kuroda}\ \emph {et~al.}(2023)\citenamefont {Kuroda},
  \citenamefont {Matsuyama}, \citenamefont {Kawasaki},\ and\ \citenamefont
  {Miyazaki}}]{kuroda2023anomalous}%
  \BibitemOpen
  \bibfield  {author} {\bibinfo {author} {\bibfnamefont {Y.}~\bibnamefont
  {Kuroda}}, \bibinfo {author} {\bibfnamefont {H.}~\bibnamefont {Matsuyama}},
  \bibinfo {author} {\bibfnamefont {T.}~\bibnamefont {Kawasaki}},\ and\
  \bibinfo {author} {\bibfnamefont {K.}~\bibnamefont {Miyazaki}},\ }\bibfield
  {title} {\bibinfo {title} {Anomalous fluctuations in homogeneous fluid phase
  of active brownian particles},\ }\href@noop {} {\bibfield  {journal}
  {\bibinfo  {journal} {Physical Review Research}\ }\textbf {\bibinfo {volume}
  {5}},\ \bibinfo {pages} {013077} (\bibinfo {year} {2023})}\BibitemShut
  {NoStop}%
\bibitem [{\citenamefont {Ramaswamy}\ \emph {et~al.}(2003)\citenamefont
  {Ramaswamy}, \citenamefont {Simha},\ and\ \citenamefont
  {Toner}}]{ramaswamy2003active}%
  \BibitemOpen
  \bibfield  {author} {\bibinfo {author} {\bibfnamefont {S.}~\bibnamefont
  {Ramaswamy}}, \bibinfo {author} {\bibfnamefont {R.~A.}\ \bibnamefont
  {Simha}},\ and\ \bibinfo {author} {\bibfnamefont {J.}~\bibnamefont {Toner}},\
  }\bibfield  {title} {\bibinfo {title} {Active nematics on a substrate: Giant
  number fluctuations and long-time tails},\ }\href@noop {} {\bibfield
  {journal} {\bibinfo  {journal} {Europhysics Letters}\ }\textbf {\bibinfo
  {volume} {62}},\ \bibinfo {pages} {196} (\bibinfo {year} {2003})}\BibitemShut
  {NoStop}%
\bibitem [{\citenamefont {Singh}\ \emph {et~al.}(2022)\citenamefont {Singh},
  \citenamefont {Pattanayak}, \citenamefont {Mishra},\ and\ \citenamefont
  {Chakrabarti}}]{singh2022effective}%
  \BibitemOpen
  \bibfield  {author} {\bibinfo {author} {\bibfnamefont {J.~P.}\ \bibnamefont
  {Singh}}, \bibinfo {author} {\bibfnamefont {S.}~\bibnamefont {Pattanayak}},
  \bibinfo {author} {\bibfnamefont {S.}~\bibnamefont {Mishra}},\ and\ \bibinfo
  {author} {\bibfnamefont {J.}~\bibnamefont {Chakrabarti}},\ }\bibfield
  {title} {\bibinfo {title} {Effective single component description of steady
  state structures of passive particles in an active bath},\ }\href@noop {}
  {\bibfield  {journal} {\bibinfo  {journal} {The Journal of Chemical Physics}\
  }\textbf {\bibinfo {volume} {156}},\ \bibinfo {pages} {214112} (\bibinfo
  {year} {2022})}\BibitemShut {NoStop}%
\bibitem [{\citenamefont {Thutupalli}\ \emph {et~al.}(2018)\citenamefont
  {Thutupalli}, \citenamefont {Geyer}, \citenamefont {Singh}, \citenamefont
  {Adhikari},\ and\ \citenamefont {Stone}}]{thutupalli2018flow}%
  \BibitemOpen
  \bibfield  {author} {\bibinfo {author} {\bibfnamefont {S.}~\bibnamefont
  {Thutupalli}}, \bibinfo {author} {\bibfnamefont {D.}~\bibnamefont {Geyer}},
  \bibinfo {author} {\bibfnamefont {R.}~\bibnamefont {Singh}}, \bibinfo
  {author} {\bibfnamefont {R.}~\bibnamefont {Adhikari}},\ and\ \bibinfo
  {author} {\bibfnamefont {H.~A.}\ \bibnamefont {Stone}},\ }\bibfield  {title}
  {\bibinfo {title} {Flow-induced phase separation of active particles is
  controlled by boundary conditions},\ }\href@noop {} {\bibfield  {journal}
  {\bibinfo  {journal} {Proceedings of the National Academy of Sciences}\
  }\textbf {\bibinfo {volume} {115}},\ \bibinfo {pages} {5403} (\bibinfo {year}
  {2018})}\BibitemShut {NoStop}%
\bibitem [{\citenamefont {Nafar~Sefiddashti}\ \emph {et~al.}(2020)\citenamefont
  {Nafar~Sefiddashti}, \citenamefont {Edwards},\ and\ \citenamefont
  {Khomami}}]{nafar2020flow}%
  \BibitemOpen
  \bibfield  {author} {\bibinfo {author} {\bibfnamefont {M.~H.}\ \bibnamefont
  {Nafar~Sefiddashti}}, \bibinfo {author} {\bibfnamefont {B.~J.}\ \bibnamefont
  {Edwards}},\ and\ \bibinfo {author} {\bibfnamefont {B.}~\bibnamefont
  {Khomami}},\ }\bibfield  {title} {\bibinfo {title} {Flow-induced phase
  separation and crystallization in entangled polyethylene solutions under
  elongational flow},\ }\href@noop {} {\bibfield  {journal} {\bibinfo
  {journal} {Macromolecules}\ }\textbf {\bibinfo {volume} {53}},\ \bibinfo
  {pages} {6432} (\bibinfo {year} {2020})}\BibitemShut {NoStop}%
\bibitem [{\citenamefont {Olmsted}(1999)}]{olmsted1999dynamics}%
  \BibitemOpen
  \bibfield  {author} {\bibinfo {author} {\bibfnamefont {P.~D.}\ \bibnamefont
  {Olmsted}},\ }\bibfield  {title} {\bibinfo {title} {Dynamics and flow-induced
  phase separation in polymeric fluids},\ }\href@noop {} {\bibfield  {journal}
  {\bibinfo  {journal} {Current opinion in colloid \& interface science}\
  }\textbf {\bibinfo {volume} {4}},\ \bibinfo {pages} {95} (\bibinfo {year}
  {1999})}\BibitemShut {NoStop}%
\bibitem [{\citenamefont {Yang}\ \emph {et~al.}(2021)\citenamefont {Yang},
  \citenamefont {Huang}, \citenamefont {Zhao},\ and\ \citenamefont
  {Zhang}}]{yang2021controlling}%
  \BibitemOpen
  \bibfield  {author} {\bibinfo {author} {\bibfnamefont {S.}~\bibnamefont
  {Yang}}, \bibinfo {author} {\bibfnamefont {M.}~\bibnamefont {Huang}},
  \bibinfo {author} {\bibfnamefont {Y.}~\bibnamefont {Zhao}},\ and\ \bibinfo
  {author} {\bibfnamefont {H.}~\bibnamefont {Zhang}},\ }\bibfield  {title}
  {\bibinfo {title} {Controlling cell motion and microscale flow with polarized
  light fields},\ }\href@noop {} {\bibfield  {journal} {\bibinfo  {journal}
  {Physical Review Letters}\ }\textbf {\bibinfo {volume} {126}},\ \bibinfo
  {pages} {058001} (\bibinfo {year} {2021})}\BibitemShut {NoStop}%
\bibitem [{\citenamefont {Durham}\ \emph {et~al.}(2009)\citenamefont {Durham},
  \citenamefont {Kessler},\ and\ \citenamefont
  {Stocker}}]{durham2009disruption}%
  \BibitemOpen
  \bibfield  {author} {\bibinfo {author} {\bibfnamefont {W.~M.}\ \bibnamefont
  {Durham}}, \bibinfo {author} {\bibfnamefont {J.~O.}\ \bibnamefont
  {Kessler}},\ and\ \bibinfo {author} {\bibfnamefont {R.}~\bibnamefont
  {Stocker}},\ }\bibfield  {title} {\bibinfo {title} {Disruption of vertical
  motility by shear triggers formation of thin phytoplankton layers},\
  }\href@noop {} {\bibfield  {journal} {\bibinfo  {journal} {Science}\ }\textbf
  {\bibinfo {volume} {323}},\ \bibinfo {pages} {1067} (\bibinfo {year}
  {2009})}\BibitemShut {NoStop}%
\bibitem [{\citenamefont {Durham}\ \emph {et~al.}(2013)\citenamefont {Durham},
  \citenamefont {Climent}, \citenamefont {Barry}, \citenamefont {De~Lillo},
  \citenamefont {Boffetta}, \citenamefont {Cencini},\ and\ \citenamefont
  {Stocker}}]{durham2013turbulence}%
  \BibitemOpen
  \bibfield  {author} {\bibinfo {author} {\bibfnamefont {W.~M.}\ \bibnamefont
  {Durham}}, \bibinfo {author} {\bibfnamefont {E.}~\bibnamefont {Climent}},
  \bibinfo {author} {\bibfnamefont {M.}~\bibnamefont {Barry}}, \bibinfo
  {author} {\bibfnamefont {F.}~\bibnamefont {De~Lillo}}, \bibinfo {author}
  {\bibfnamefont {G.}~\bibnamefont {Boffetta}}, \bibinfo {author}
  {\bibfnamefont {M.}~\bibnamefont {Cencini}},\ and\ \bibinfo {author}
  {\bibfnamefont {R.}~\bibnamefont {Stocker}},\ }\bibfield  {title} {\bibinfo
  {title} {Turbulence drives microscale patches of motile phytoplankton},\
  }\href@noop {} {\bibfield  {journal} {\bibinfo  {journal} {Nature
  communications}\ }\textbf {\bibinfo {volume} {4}},\ \bibinfo {pages} {2148}
  (\bibinfo {year} {2013})}\BibitemShut {NoStop}%
\end{thebibliography}%

\end{document}